\documentclass[twocolumn,prb,showpacs,superscriptaddress,floatfix]{revtex4}

\usepackage{graphicx}
\usepackage{bm}
\usepackage{epstopdf}
\usepackage{amsmath,amsfonts,paralist}

\newcommand{\BEQ}{\begin{equation}}
\newcommand{\EEQ}{\end{equation}}
\newcommand{\BEA}{\begin{eqnarray}}
\newcommand{\EEA}{\end{eqnarray}}

\newcommand{\p}{\partial}
\newcommand{\nn}{\nonumber }

\makeatletter
\newcommand\figcaption{\def\@captype{figure}\caption}
\makeatother

\begin{document}
\title{The random Blume-Capel model on cubic lattice:  first order inverse freezing in a 3D
spin-glass system}
\date{\today}

\author{L. Leuzzi}
\email{luca.leuzzi@cnr.it}
\affiliation{IPCF-CNR, UOS Roma,  P.le Aldo Moro 2, I-00185 Roma, Italy}
\affiliation{Dipartimento di Fisica,  Universit\`a "Sapienza",
  P.le Aldo Moro 2, I-00185 Roma, Italy}

\author{M. Paoluzzi}
\affiliation{IPCF-CNR, UOS Roma,
  P.le Aldo Moro 2, I-00185 Roma, Italy}
\affiliation{Dipartimento di Fisica,  Universit\`a di Roma 3,
Via della Vasca Navale 84, I-00146 Roma, Italy}

\author{A. Crisanti}
\affiliation{Dipartimento di Fisica, Universit\`a "Sapienza",
  P.le Aldo Moro 2, I-00185 Roma, Italy}

\pacs{}

\begin{abstract}
We present a numerical study of the Blume-Capel model with quenched
disorder in 3D. The phase diagram is characterized by
spin-glass/paramagnet phase transitions of both first and second order
in the thermodynamic sense. Numerical simulations are performed using
the Exchange-Monte Carlo algorithm, providing clear evidence for
inverse freezing.  The main features at criticality and in the phase
coexistence region are investigated.  The whole inverse freezing
transition appears to be first order.  The second order transition
appears to be in the same universality class of the Edwards-Anderson
model. The nature of the spin-glass phase is analyzed by means of the
finite size scaling behavior of the overlap distribution functions and
the four-spins real-space correlation functions. Evidence for a
replica symmetry breaking-like organization of states is provided.

\end{abstract}

\maketitle

\section{Introduction}

\label{sec:intro}

 The so-called {\em inverse transition} (IT) is a reversible
transformation occurring between phases whose entropic contents and
symmetries are in the inverse order relation relatively to standard
transitions. The case - already hypothesized by Tammann more than a
century ago \cite{Tammann} - of ``ordering in disorder'' taking place
in a crystal solid that liquefies on cooling, is generally termed {\em
inverse melting}.  The IT phenomenon also includes the transformation
involving amorphous solid phases - rather than crystal - as that of a
liquid vitrifying upon heating. In this case the term {\em inverse
freezing} is somewhat used in the literature: both phases are
disordered but the fluid appears to be the one with least entropic
content.  The reason for these counter intuitive phenomena is that a
phase usually present only at high temperature happens to exist also
in peculiar patterns such that its entropy is actually less than the
one of the phase normally considered the most ordered one.
\\ \indent Inverse transitions in their most generic meaning (i.e.,
both thermodynamic or occurring by means of kinetic arrest) have been
detected in the last years in a number of different materials and
between phases of rather different nature.  The first example was the
transition between liquid and crystal phases of helium isotopes He$^3$
and He$^4$ at low temperature.\cite{Wilks87}  A more complex and
recent example is the polymer poly(4-methylpentene-1) - P4MP1 - in
which a crystal polymer melts as the temperature is decreased, or the
pressure increased.  By means of exhaustive measurements by
Differential Scanning Calorimetry (DSC) and X-ray diffraction the
phase diagram of P4MP1 has been experimentally determined by the group
of Rastogi, \cite{RHKMM99,Greer00,vRRMM04} showing evidence for both
an equilibrium inverse melting, between a crystal phase (tetragonal or
hexagonal, depending on the pressure) and a fluid phase, and a
non-equilibrium IT between the hexagonal crystal and a glassy
phase. Another extensively studied instance is a molecular solution in
water, composed by $\alpha$-cyclodextrine ($\alpha$CD) and
4-methylpyridine (4MP) mixed in given molecular ratios, investigated
by means of neutron scattering, X-ray diffraction, DSC and rheometric
measurements.
\cite{Plazanet04,Tombari05,Plazanet06,Plazanet06b,Angelini07,
Ferrari07,Angelini08,Angelini08b,Plazanet09,Angelini09} The "solid" is
in this case a sol-gel porous system formed by an ordered network of
molecules of $\alpha$CD-water-4MP filled with liquid 4MP, melting down
decreasing temperature at constant $\alpha$CD concentration.
Eventually, another important polymeric example is methyl-cellulose
solution in water, undergoing a reversible inverse sol-gel transition.
\cite{CACPS97, Hirrien98} For such system, a careful analysis of the
behavior of the microscopic components across the transition has been
performed in literature \cite{Haque93} and, therefore, it turns out to
be particularly important for the modelization proposed in the present
work, as we will see in the following.
\\ \indent Apart from polymeric and macromolecular substances, in the
last years ITs showed up in many other different contexts.  Inverse
melting from an ordered lattice to a disordered vortex phase takes
place, e.g., for the magnetic flux lines in a high temperature
superconductor. \cite{Avraham01} A gas of atoms at zero temperature
passes from superfluid to insulator as the lattice potential depth is
increased. \cite{Greiner02} Furthermore, in the framework of
nanosystems, the reversible transition of an isotropic liquid into an
ordered cubic phase upon heating has been detected experimentally in
ferromagnetic systems of gold nanoparticles. \cite{Donnio07,Donnio10}
%
%%%%
\\ \indent In this work we stick to a definition of IT as the one put
forward by Tammann: \cite{Tammann} a reversible transition in
temperature at fixed pressure - or generally speaking, at a fixed
parameter {\em externally} tuning the interaction strength, such as
concentration, chemical potential or magnetic field - from a solid
high temperature phase to an isotropic fluid (or a paramagnet, for
magnetic systems) low temperature phase.  Generalizing to
non-equilibrium systems one might address as IT also those cases in
which the isotropic fluid is dynamically arrested into a glassy
state. This occurs, e.g., for the crystal-glass transition in the
cited P4MP1 as pressure is not too large \cite{RHKMM99} or in
molecular dynamics simulations and mode-coupling computations of
attractive colloidal glasses. \cite{Zaccarelli02,Zaccarelli04}
\\
\indent
In this definition IT is not an exact synonym of reentrance.  Indeed,
though a reentrance in the transition line is a common feature in ITs,
this is not always present, as, e.g., in the case of $\alpha$CD
\cite{Plazanet04,Tombari05,Angelini07} or methyl-cellulose
\cite{CACPS97} solutions for which no high temperature fluid phase has
been detected. Moreover, not all re-entrances can be seen as
signatures of an IT to a completely disordered isotropic phase.  In
liquid crystals, ultra-thin films and other materials, phases with
different kind of symmetry can, actually, be found that are separated
by reentrant isobaric transition lines in temperature - cf., e.g.,
Refs. [\onlinecite{Cladis75,Cladis77,Ozbek02,Portmann03,Srivastava07}] -
but no melting occurs {\em strictu sensu}.  Also re-entrances between
dynamically arrested states, aperiodic structures or amorphous solids
of qualitatively similar nature, like liquid-liquid pairs
\cite{Jaffar97, Bagchi06} are not considered as IT, since an a-priori
order relationship between the entropic content of the two phases is
not established and it cannot be claimed what is inverse and what is
"standard". For the same reasons also re-entrances between spin-glass
(supposed at {\em equilibrium}, that is, considered as a thermodynamic
phase) and ferromagnetic phases - as, e.g., in
Refs. [\onlinecite{Verbeek78,Yeshurun80}] - hardly fall into the IT
category.  Eventually, re-entrances in parameters other than
temperature are also not taken into account as inverse
melting/freezing transitions.
\\ \indent A thorough explanation of the fundamental mechanisms
leading to the IT would need of a microscopic analysis of the single
components behavior and their mutual interactions as temperature
changes across the critical point.  Due to the complexity of the
structure of polymeric chains and macromolecules involved in such
transformations, a clear-cut picture of the state of the single
components is often not available. For the case of the above mentioned
methyl-cellulose, Haque and Morris\cite{Haque93} proposed that chains
exist in solution as folded bundles in which hydrophobic methyl groups
are packed.  As the temperature is raised, the bundles unfold,
exposing methyl groups to water molecules and, thus, causing a large
increase in volume and the formation of hydrophobic links eventually
leading to a gel. The polymers in the folded state are thus inactive
(or far less interacting than those in the unfolded state) but also
yield a smaller entropic contribution than the unfolded ones.  As the
chains start to unfold because of thermal noise they change to an
interacting state thus enforcing bonds with other chains and
condensing in a gel.
\\ \indent Theoretical modeling for IT is starting to develop but is
 still on its first, often uncorrelated, steps and consists, at the
 better, in heuristic reproductions of the phenomenon.\cite{SDTJPC01,
 SDBBPC03, Feeney,SSPRL04, SS05, Prestipino, CLPRL05, Sellitto06}
 Looking, in particular, at the transition between an amorphous
 'frozen' phase and a fluid (i.e., paramagnet), recently spin-glass
 models with spin-$1$ variables have turned out to effectively
 represent systems in which the transformation is driven by entropic
 effects. In these cases inverse freezing has been studied in the
 mean-field approximation.\cite{CLPRL05, Sellitto06}
\\ \indent We also mention that with the help of this class of models,
the connection between entropy driven phase reentrance and shear {\em
thickening} can also be tackled\cite{SKPRL05} and, furthermore, a
generalization of the spin-$1$ variable to a composition of ``fast''
and ``slow'' variables \footnote{E.g., setting $S=\sigma n$, with
$n=0,1$ fast and $\sigma = \pm 1$ slow} coupled to two different
thermal baths allows for studying anomalous latent heat in out of
equilibrium transitions.\cite{APPRL06}
 \\ \indent In the present work we will consider the Blume-Capel (BC)
 model\cite{CapelPhys66} with quenched disorder: a spin-glass model on
 a 3D cubic lattice with bosonic spin$-1$ variables ($s_i= -1,0,+1$).
 Under the assumption that the interplay between inactive and
 interactive states of a microscopic component is at the ground of the
 eventual IT, bosonic spins can approximate the folded/unfolded
 conformation, $S=0$ representing the inactive state, $S=\pm 1$ the
 interactive one, cf. Ref. [\onlinecite{SS05}] for a more
 comprehensive discussion.  We will focus on the random version of the
 BC model introduced by Ghatak and Sherrington \cite{GSJPC77} (GS) in
 order to study the effects of the crystal-field in a spin glass -
 e.g., (Ti$_{1-x}$V$_x$)$_2$O$_3$ displays anisotropic spin glass
 behavior in function of $x$.\cite{Dumas} The mean-field solution in
 the Full Replica Symmetry Breaking (RSB)
 scheme\cite{CLPRL05,CLPRL02,CLPRB04,LPM} predicts a phase diagram
 with a second order transition line between spin-glass (SG) and
 paramagnetic (PM) phase ending in a tricritical point where a first
 order phase transition line starts and a phase coexistence region
 appears.
 \footnote{We stress that the transition is first order in the
 thermodynamic sense, with latent heat and is not related to the
 so-called random first order transition occurring in mean-field
 models for structural glasses.}  Furthermore, the first order
 transition is characterized by the phenomenon of
 IT:\cite{CLPRL05,LPM} the low temperature phase is PM with a lower
 entropy than the SG phase and the transition line develops a
 reentrance.
\\ \indent In the original (ferromagnetic) BC
models\cite{CapelPhys66,BEGPRA71} however, no IT was observed in the
mean-field approximation, nor in finite dimension
studies.\cite{Saul74,Berker76,Jain80,Hasenbusch10} and in presence of
quenched disorder a recent study on a 3D hierarchical lattice by means
of renormalization group theory in position space \cite{Ozcelik08}
provides no evidence for a low temperature tricritical point or a
PM/SG reentrance, contrarily to what is predicted by mean-field
theory.
 \\ \indent Moreover, we found in the literature only one finite
dimensional system with quenched disorder undergoing a standard first
order phase transition in finite dimension: the 4-Potts glass studied
by Fernandez {\em et al.} in Ref. [\onlinecite{Fernandez08}]. In that
work a first claim has been made that first order phase transition
exists in 3D systems also in presence of quenched disorder, though the
randomness tends to strongly smoothen the transition into a second
order one.  This transition is driven by the temperature and by the
degree of dilution of the Potts glass bonds. Though in numerical
simulations changing, e.g., the pressure, the bond dilution, or even
the relative probabilities of the random bond values, cf., e.g.,
Ref. [\onlinecite{Toldin09}], is technically equivalent, the latter
are complicated to control in a real experiment and require the
preparation of several samples with different microscopic properties.
The study of a conceptually simpler model, satisfactorily approachable
with standard simulation techniques, might help in validating the
assessment of the existence of first order phase transitions in random
systems.
\\ \indent Motivated by the above considerations we have, thus,
studied the existence of inverse freezing in the 3D disordered BC
model with nearest-neighbor interactions and the nature of the
``frozen'' (or, rather, ``blocked'') phase.  We present hereafter the
results of our investigation by means of Monte Carlo
numerical simulations. Some results about the critical behavior have
already appeared in a recent letter.\cite{Paoluzzi10}
\\ \indent In the present manuscript we first introduce the model, in
Sec. II, and in Secs. III and IV we define the numerical techniques
employed to study continuous and discontinuous phase transitions in
finite size (FS) systems.  In Sec. V we recall the Exchange Monte
Carlo method, else called Parallel Tempering
(PT).\cite{Hukushima96,Marinari98} In Sec. VI, we present our results
about the phase diagram of the model and its critical behavior both
along the continuous transition and in the coexistence region related
to the first order transition. The main features of the organization
of states in the SG phase in finite dimension (i.e., below the upper
critical dimension for our model) is studied in Sec. VII, where we
perform a systematic study of the properties of the overlap
distribution functions and of the four-spins correlation functions in
space. Finally, Sec.  VIII reports our conclusions.

%{\em The  model.}
\section{Model and order parameters}
We consider the following Hamiltonian
\BEQ
\mathcal{H}_J[s]=-\sum_{({i}{j})}
J_{{i}{j}}s_{{i}}s_{{j}}
+D\sum_{{i}}s^2_{{i}} \EEQ
where $({i}{j})$ indicates ordered couples of
nearest-neighbor sites, and $s_{{i}}= -1,0,+1$ are spin$-1$
variables lying on a cubic lattice of size $N=L^3$ with Periodic
Boundary Condition (PBC).  Random couplings
$J_{{i}{j}}$ are independent identically
distributed as
 \BEQ
 P(J_{{i}{j}})=\frac{1}{2}
\delta (J_{{i}{j}}-1)
+\frac{1}{2}\delta(J_{{i}{j}}+1)
 \EEQ
The field $D$ is usually called crystal-field and it plays the role of a chemical potential for the 
empty sites $s=0$. We will, therefore, refer to $D$ invariably as chemical potential or crystal field in the following.
 We simulate two real replicas of the system and define their site and
link overlaps, i.e., the order parameters usually characterizing the
SG transition, as
 \BEA
 q_{s}^{(J)}&\equiv&
\frac{1}{N}\sum_{{i}}s^{(1)}_{{i}}
s^{(2)}_{{i}}
\\
q_{l}^{(J)}&\equiv&
\frac{1}{3N  }\sum_{({j}
{k})}
 s^{(1)}_{{j}}s^{(1)}_{{k}}
s^{(2)}_{{j}}s^{(2)}_{{k}}
\EEA
 where $3$ is the
 dimension of the space. If a thermodynamic
  phase transition occurs, with latent heat, the most significant
  order parameter that drives the transition is the density $\rho$ of
  magnetically active ($|s_i|=1$) sites:
\BEQ
\rho^{(J)}=\frac{1}{N}\sum_{{i}} s^2_{{i}}
\EEQ
The apex $J$ recalls us that the values of the parameters depend on
the particular realization of disorder ($\{J_{{i}{j}}\}$).
Useful information about the equilibrium properties of
the system can be obtained from the knowledge of the following probability
distribution functions (pdf)
\BEA
P(q_s)&\equiv&\overline{P_J(q_s)} = \overline{\left\langle \delta
\left(q_s- q_{s}^{(J)} \right)
\right\rangle}
\label{f:pqs}
\\
P(q_l)&\equiv& \overline{P_J(q_l)}
\label{f:pql}
=\overline{\left\langle \delta \left(q_l - q_{l}^{(J)} \right)
\right\rangle}
\\
P(\rho)&\equiv& \overline{P_J(\rho)} =
\overline{\left\langle \delta \left(\rho- \rho^{(J)}\right) \right\rangle}
\label{f:prho}
  \EEA
where ${\overline{\phantom{|}\ldots \phantom{|}}}$ denotes the average
over quenched disorder and $\langle {\phantom{|}\ldots \phantom{|}}
\rangle$ the thermal average.  Though the density probability
distribution is known to be self-averaging ($\lim_{N\to \infty
}P_{J,N}(\rho)={\overline{P_J(\rho)}}$), this does not hold for the
overlap distributions $P_J(q_{s,l})$,\cite{MPV86} for which
\footnote{In the present work the probability distributions always
depend on the size $N$, or $L$. In order not to make the notation too
heavy, we will explicitate it only when needed.}
\BEA
P(q_{s,l})&\equiv&
\overline{P_J(q_{s,l})}\neq \lim_{N\to\infty} P_{J,N}(q_{s,l})
\EEA
\\ \indent Through the study of the pdfs, as function of the external
thermodynamic parameters, we can identify the PM /SG transition and
discriminate between first and second order phase transitions.  As an
instance, if a first order phase transition takes place, the density
pdf $P(\rho)$ displays a double peak due to the coexistence of the PM
and SG phases.  Moreover, by means of the overlap pdfs we can
investigate the nature of the SG phase.

\section{Finite Size Scaling for Continuous Transitions}
\label{sec:uniFSS}
In order to infer the details of the critical behavior from numerical
simulations of finite size systems, a fundamental quantity (in zero
external magnetic field) is
\BEA
C_4({\bm r})
\equiv
\frac{1}{N}
\sum_{\bm{s}}
{\overline{\langle
s_{\bm s}s_{\bm s+\bm r}
\rangle^2}}
\label{f:C4def}
\EEA
with $\bm{r} = (r_x,r_y,r_z)$.
In terms of space-dependent overlaps,
$q_{\bm r}=s^{(1)}_{\bm r} s^{(2)}_{\bm r},$ $C_4$ can be written as
\BEA C_4({\bm r}) &\equiv&\frac{1}{N} \sum_{{\bm p}} {\overline{\langle
q_{{\bm p}} q_{{\bm p}+{\bm r}} \rangle_{12}}}
\\
\nn
&&=\frac{1}{N} \sum_{{\bm p}} {\overline{\langle
s^{(1)}_{{\bm p}} s^{(1)}_{{\bm p}+{\bm r}} \rangle_{1}\langle
s^{(2)}_{{\bm p}} s^{(2)}_{{\bm p}+{\bm r}} \rangle_{2}}} 
\EEA
where $\langle \ldots
\rangle_{12}$ stays equivalently for
the thermal average $\langle \langle \ldots
\rangle_1\rangle_2$ or $\langle \langle \ldots
\rangle_2\rangle_1$ over the two replicas independently.
This is the four-spins correlation function, and the information it
provides can be exploited in different ways to identify the existence
of a second order phase transition for finite size systems and probe the 
thermodynamic behavior in the SG phase.
\\
\indent
A conventional  way to identify a second
order phase transition is to look at the behavior of a correlation
length-like scaling function defined as
\BEQ
\xi_c^2\equiv\frac{\int dr\, r^2 C_4(r)}{\int dr\, C_4({\bm r})}
=\left.\frac{\p \log{\hat{C}_4 ({\bm k})}}{\p k^2} \right|_{k^2=0}
\EEQ
where
\BEQ
 \nn
\hat{C}_4({\bm k})=\frac{1}{(2\pi)^3}\int d^3 r
\,e^{-i{\bm k}\cdot{\bm r}}\,C_4({\bm r}) \ .
\EEQ
On a 3D cubic lattice, the above defined correlation length  becomes:
\cite{Caracciolo93}
\BEQ
\xi_c^2=\frac{1}{4\sin^2{\frac{k_1}{2}}}
\left(\frac{\hat{C}_4(\bm 0)}{\hat{C}_4({\bm k_1})}-1\right)
\label{f:xic}
\EEQ
 where $k_1=|{\bm k_1}|$,
${\bm k_1}\equiv(2\pi/L,0,0)$ is the minimum
wave-vector of the lattice and $\bm 0 = (0,0,0)$.
 In the thermodynamic limit, a second order transition is
characterized by a diverging correlation length, at critical
temperature $T_c$, whose Finite Size Scaling (FSS) behavior is the same as in
Eq. (\ref{f:xic}). \cite{palassini,BCPRB00}
\\
\indent
 Another relevant
observable is the SG susceptibility
\BEQ \chi_{SG}\equiv L^3
\overline{\langle q^2 \rangle}=L^3\hat{C}_4(\bm 0)
\label{def:chisg}
\EEQ
diverging at the
PM/SG transition as $L \to\infty$. Because of FS, though, $\xi_c$ and
$\chi_{SG}$ cannot diverge in numerical simulations, although inside
the critical region a remarkable property of the critical phenomena
survives: scale invariance. Indeed, we can define a FS ``critical''
temperature $T_c^L$, function of the size $L$, as the temperature at
which the above mentioned observables do not
(or only slightly) depend on the size:
\BEA
\frac{\xi_c}{L}&=&\bar \xi_c\left(\frac{\xi_c}{L}\right)
= \bar\xi(L^{1/\nu}(T-T_c^L))
\\
\chi_{SG} L^{\eta-2}&=&\bar \chi\left(\frac{\xi_c}{L}\right)
= \bar \chi(L^{1/\nu}(T-T_c^L))
\EEA
The values $T_c^L$ at which $\xi_c/L$ at different $L$ cross each
other are the FS respective of the critical temperature. The latter
can, thus, be estimated by FSS in the $L\to \infty$ limit. 
\\
\indent
 A further
size independent observable at criticality is the so-called Binder
parameter:
\BEQ
g(L,T)=\frac{1}{2}\left(3-\frac{q_4}{(q_2)^2}\right)
\label{f:binder}\EEQ
with $q_n
\equiv \overline{\langle (q_s^{(J)})^n \rangle}$.  It measures the
deviation of $P(q)$ from a Gaussian distribution as the SG phase is
approached. Since $q_4$ and $q_2^2$ scale with $L$ in the same way,
$g$ does not depend on $L$ at $T_c$.

\subsection{Quotient method}
 Denoting by $O(T,L)$ a generic observable diverging at critical
temperature $T_c$ as $L\to \infty$, and considering two sizes $L$,
$L'$ whose scale ratio is $s=L'/L$, we can look at the scaling of quotient 
 \BEA \label{scaling}
\frac{O(T,sL)}{O(T,L)}=F_O\left(\frac{\xi_c(L,T)}{L},s\right)
+\mathcal{O}(\xi_c^{-\omega},L^{-\omega}) \EEA
where $F_O$ is a universal FSS function and $\omega$ is the power of
the subleading FS corrections.  Through the scaling {\em Ansatz},
(\ref{scaling}) one, thus, introduces a class of universal functions
$F_{O}$ that are size-independent in the critical region. Given two
observables $O$ and $R$ displaying scale invariance, this allows for
plotting $F_O$ vs $F_R$ for several values of $L$: if the data
collapse in a universal scaling function, the scaling Ansatz is
verified and FSS methods can be trusted to evaluate the critical
exponents. We will analyze in the present manuscript the behavior of
$F_\xi$, $F_{\chi_{\rm SG}}$ and $F_g$.
\\
\indent
In order to estimate the critical exponents we can, thus,
 use the so-called {\em quotient method}, \cite{BCPRB00} based on the
 observation that at  $T_c^L\equiv T_c^*$, the correlation
 lengths in systems of sizes $L$ and $sL$ (in $L$ units) are equal:
\BEQ
\frac{s\,\xi_c(T_c^*,L)}{  \xi_c(T_c^*,sL)}=1.
\EEQ
\\
Indent
For an observable $O$ scaling as $t^{x_O}$ ($t=T/T_c-1$) in the
thermodynamic limit, it holds:
 \BEQ
s^{\frac{x_O}{\nu}}=\frac{O(T_c^*,sL)}{O(T_c^*,L)}+\mathcal{O}(L^{-\omega})
\EEQ
where the dependence through $\xi_c^{-\omega}$ in the correction term is
neglected because, in the critical region, $\xi_c\gg L$.  For a SG we
can obtain the exponents $\nu$ and $\eta$ by means of the FSS of the
quotients of $\p_\beta\xi$ and $\chi_{SG}$, scaling,
respectively, with exponents
\BEA
x_{\p_\beta\xi_c}&=&1+\nu
\nn
\\
\nn
x_{\chi_{SG}}&=&(2-\eta)\nu \; .
\EEA
These relations hold if the Ansatz (\ref{scaling}) is verified,
\cite{Caracciolo93} i.e., if $F_{O}$ is a size-independent
scaling function for several values of $L$ and $sL$.

\section{Characterization of a first order transition}
\label{sec:fo}
Besides the second order transition, the random BC  mean-field model also shows
a  tricritical point beyond which a true first order phase
transition with non-zero latent heat occurs and from
which a region of coexistence of PM and SG phase departs.
\cite{CLPRL05}    The
slope of a first order line is given by the Clausius-Clapeyron
equation that, for our model reads 
\BEQ
\frac{dD}{dT}=\frac{s_{PM}-s_{SG}}{\rho_{PM}-\rho_{SG}}=\frac{\Delta
s}{\Delta\rho}
\label{eq:CC}
 \EEQ
where $D$ plays the role of the
pressure. The equilibrium transition
line changes slope in a point where the entropy of the two coexistence
phase is equal $\Delta s =0$ (Kauzmann locus \cite{SDTJPC01,Kauz48}).
\\ \indent In order to tackle the identification of a first order
phase transition in a 3D finite size system from numerical simulation
data we sketch in the following four methods to estimate critical (and
spinodal) points.
%\\
%\indent
\subsection{``Equal weight'' estimate}
The transition takes place at the point at which the configurations
belonging to the SG phase and those belonging to the PM phase have the
same statistical weight, i.e., they yield identical contribution to
the partition function of the single pure phase. Else said, the free
energies of the two coexisting phases are equal.  In this statistical
mechanical framework, the FS transition line $D_c(L,T)$ can be
obtained as the locus of points where the two phases are equiprobable,
i.e., the areas of the two peaks are equal: \cite{Hill}
\BEQ
\int_{0}^{\rho_0} d\rho\,P(\rho)=\int_{\rho_0}^1
 d\rho\, P(\rho)
\label{f:fo_ew}
\EEQ
where $\rho_0\in [\rho_{PM}:\rho_{SG}]$ such that $P(\rho_0)=0$ (or
 minimal next to the tricritical point).  A way to numerically
 determine the transition point is, thus, to compare the areas under
 the peaks of the distributions, cf. Eq. (\ref{f:prho}).

\subsection{Maxwell ``Equal distance'' estimate}
There are other two methods to determine a first order transition in
finite systems, based on the Maxwell construction.  If we are in the
coexistence region the curve $D(\rho)$ for the system of size $L$ will
display a sort of plateau around some $D=D^\star\simeq D^L_c$: in a
very small interval of $D$ the density changes very rapidly. In
the case of a pure state, instead, the $D(\rho)$ curve has a far
smoother behavior. In order to estimate the critical point we need to
extrapolate the behavior of $D(\rho)$ for the pure phases inside the
region of coexistence. To this aim we perform two fits exclusively
based on the points outside the spinodal points: for $D<D_{\rm SG}^{\rm sp}$
for the SG phase ($D_{\rm SG}(\rho)$) and $D>D_{\rm PM}^{\rm sp}$ for the
PM phase  ($D_{\rm PM}(\rho)$).
We will call $\rho_{\rm PM,SG}(D)$
 the inverse of the curves $D_{\rm PM,SG}(\rho)$ extrapolated from
the data points pertaining to the pure PM and SG phases,
respectively.  The curve $\rho(D)$ will denote the inverse of $D(\rho)$.
\\
\indent
In this way we can make  a
Maxwell-like construction in the $(\rho$, $D)$ plane at a given
temperature and determine the value of $D_c$ as the one whose
corresponding $\rho_{\rm ed}$ value along the $D(\rho)$ FS curve is equally
distant from both $\rho_{\rm PM}(D_c)$ and $\rho_{\rm SG}(D_c)$,
cf. Fig. \ref{fig:disegno}:
\BEQ
\rho_{\rm ed}(D_c) = \frac{1}{2}\left[\rho_{\rm PM}(D_c) + \rho_{\rm
SG}(D_c)\right]
\EEQ

\subsection{Maxwell ``Equal area'' estimate}
\begin{figure}[t!]
\includegraphics[width=.95\columnwidth]{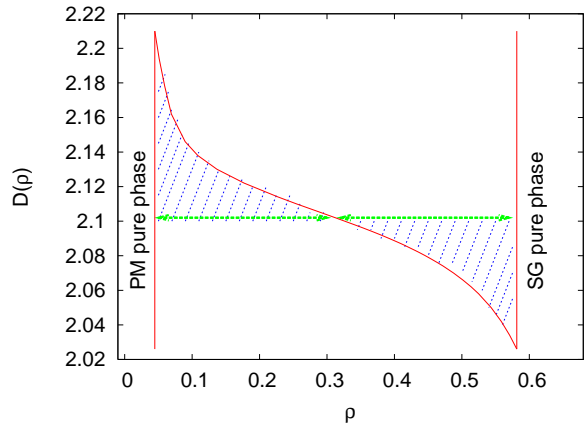}
\caption{ Graphical sketch of the equal distance (dashed/green arrows)
and equal area methods (dotted/blue areas) for FS systems undergoing
first order phase transition for a given instance ($L=6, T=0.5$). }
\label{fig:disegno}
\end{figure}

 Alternatively we can determine $D_c$ as the value at which,
 cf. Fig. \ref{fig:disegno}
\BEQ \int_{D_{\rm SG}}^{D_c} \rho_{\rm PM}(D) dD + \int_{D_c}^{D_{\rm
PM}} \rho_{\rm SG}(D) dD = \int_{D_{\rm SG}}^{D_{\rm PM}} \rho(D) dD
\EEQ
where $D_{\rm SG}$ and $D_{\rm PM}$ are arbitrary, provided that they
pertain to the relative pure phases.

\subsection{Symmetric distribution estimate}
Defining the skewness of the density probability distribution as
 \BEQ
 \zeta(\langle\rho\rangle)=\frac{\langle (\rho-\langle\rho\rangle)^3 \rangle}{\langle
 (\rho-\langle\rho\rangle)^2\rangle^{3/2}}
\label{f:skew}
\EEQ in Ref. [\onlinecite{Katzgraber_skew}] the critical point was
estimated as the point at which the double peaked distribution is
symmetric. Since the skewness of $P(\rho)$ can be precisely computed
this estimate does not suffer, e.g., of the arbitrariness of the
choice of $\rho_0$. In the thermodynamic limit, indeed, in the phase
coexistence region both peaks of $P(\rho)$ should be Dirac delta
distributions and equal weight would be equivalent to a
symmetric bimodal distribution.  We will show the outcome of this
analysis in our system for different cases and compare it with the
previous ones.

\section{Exchange Monte Carlo algorithm  in T and D}
\begin{figure}[b!]
\includegraphics[width=.95\columnwidth]{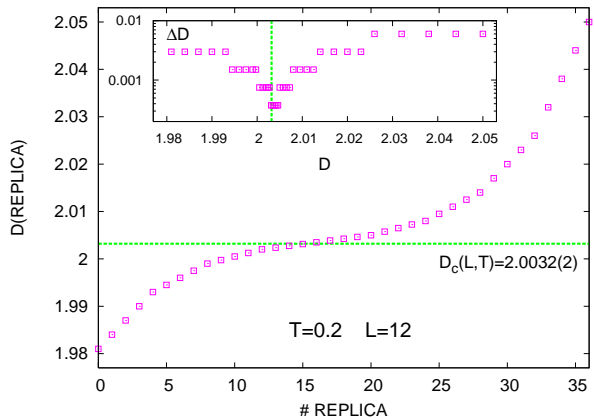}
\caption{Values of the chemical potential $D$ for the
replicas in the PT simulation exchanging systems at different $D$. The
parameters refer to the simulated $L=12$ system at $T=0.2$. The dashed
(green) line is the estimate of the FS critical value $D_c(L,T)$
estimated by means of the equal weight method (cf. Sec. \ref{sec:fo}).
Inset: Chemical potential intervals $\Delta D$ vs. $D$ in log scale
for the same instance.}
\label{fig:DD_vsD}
\end{figure}

\begin{figure}[t!]
\includegraphics[height=.95\columnwidth, angle=270]{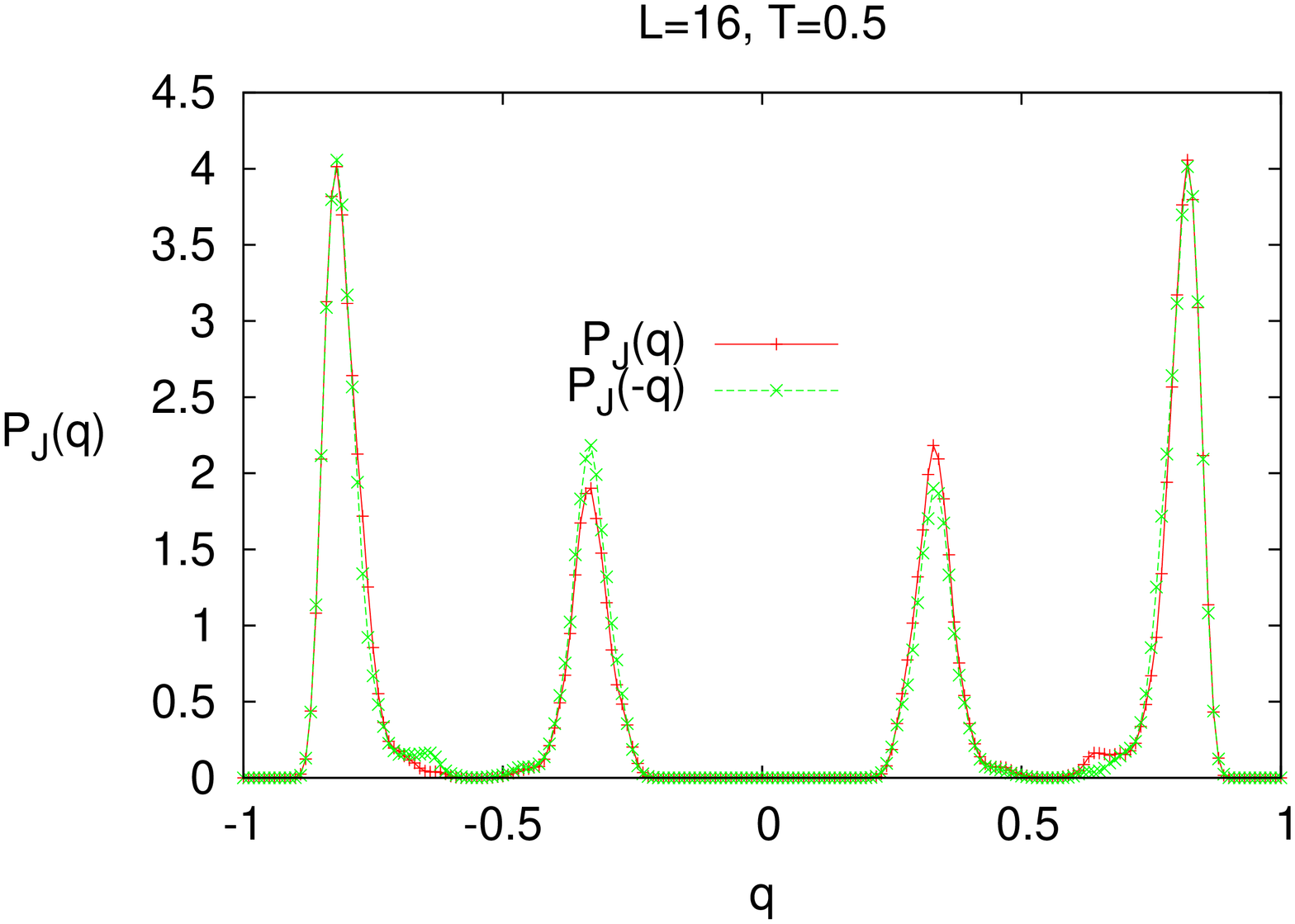}
\\
\includegraphics[height=.95\columnwidth,angle=270]{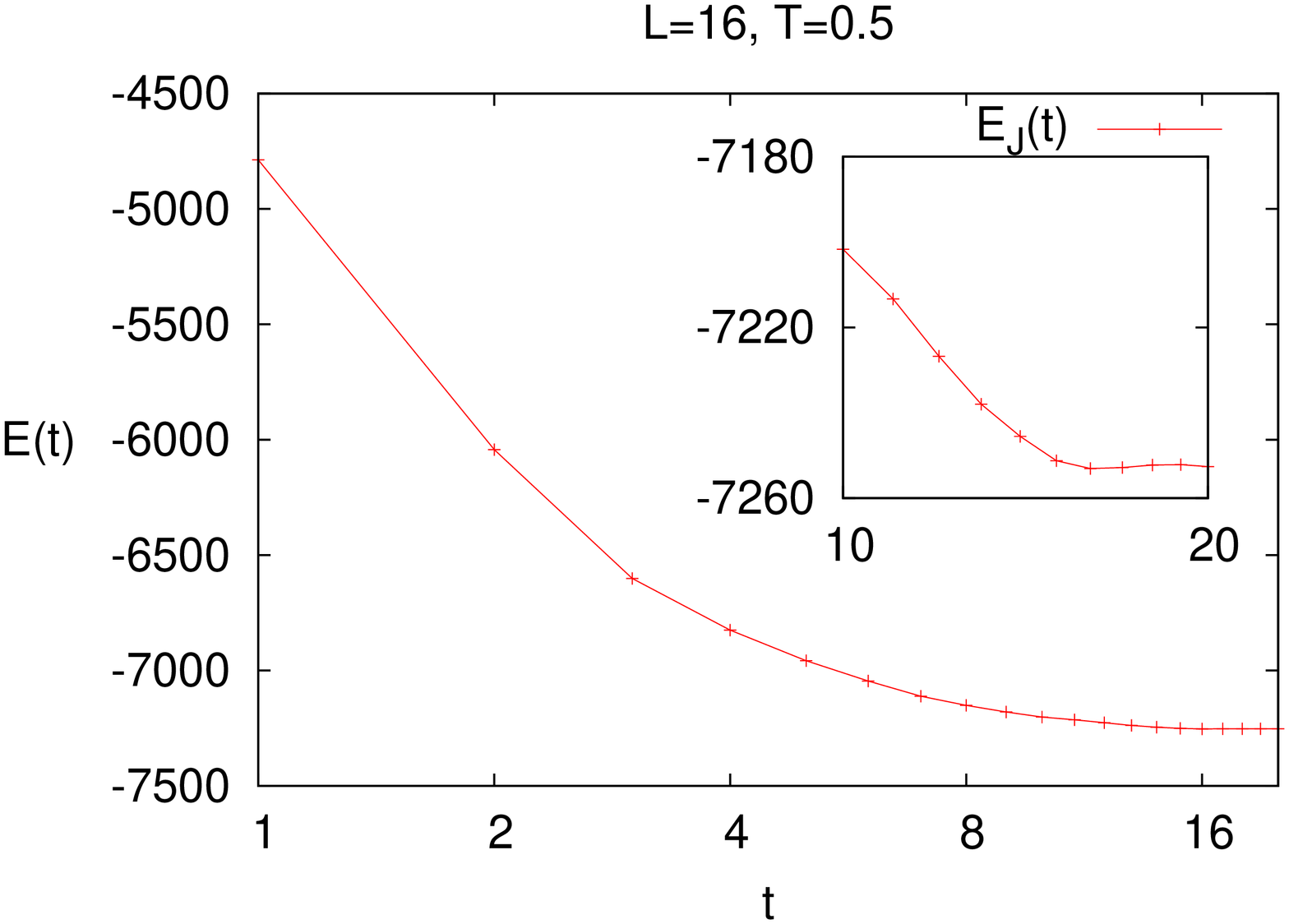}
\caption{Instances of thermalization checks. Top: $P_J(q_s)$ and
$P_J(-q_s)$ for an arbitrary sample at $D=0$, $T=0.5$, $L=16$. Bottom: average
energy versus time (in MCS) in log scale, $t=\log (\# MCS)/\log 2$.}
\label{fig:thermalization}
\end{figure}

We have simulated our spin-1 model using the parallel tempering (PT)
 algorithm, replicating several copies of the system both at different
 temperatures and at different values of the external field $D$.  For
 the PT in temperature, the swap probability of two copies at
 temperature $T$ and $T+\Delta T$ is:
\BEQ
P_{\rm swap}(\Delta \beta)=
\min\left[1,\exp\{\Delta\beta\Delta\mathcal{H}\}\right],
\EEQ
with $\Delta\beta = 1/(T+\Delta T)- 1/T$; whereas   the  swap
probability in the chemical potential reads
\BEQ
P_{\rm swap}(\Delta D)=\min\left[1,\exp\{\beta\Delta
D\Delta\rho\}\right]
 \EEQ
We used the latter implementation in trying to identify the reentrance
of the transition line in the $T,D$ plane.  Since, however, the
transition turns out to be first order in the whole region of inverse
freezing, the PT algorithm must be handled with caution.  Indeed, at
the transition $\Delta \rho$ is discontinuous implying the vanishing
of $P_{\rm swap}(\Delta D)$ around the critical point for a given fix
probe $\Delta D$.  In order to overcome this problem we have used a
varying $\Delta D$, smaller in the candidate coexistence region and
larger and larger outside.  An instance of this kind choice is
represented in Fig. \ref{fig:DD_vsD}.
\\ \indent For very large sizes, though, this would require a too
precise {\em a-priori} knowledge of the transition lines and the
method could not be applied with a reasonable success. On
the other hand, the FSS effects appear to be almost nonexistent at the
first order phase transition, so that probes at larger sizes were
actually not necessary.  In Tabs. \ref{tab:param}, \ref{tab:param2} we
report our simulation parameters for PT in temperature and in crystal field, respectively.
\begin{table}[!b]
\begin{tabular}{| c | l || c | c | c | c | c | c | c |} \hline\hline
$D\downarrow$   &$L\rightarrow$ &       6       &       8       &       10      &       12      &       16      &       20              &       24                      \\  \hline\hline
0.0     &$T_{in}$                               &       $0.6$   &       $0.6$   &       $0.7$   &       $0.7$   &       $0.85$  &       $0.9$   &       $1.0$           \\
                &$N_T$                          &       $37$            &       $37$            &       $33$            &       $33$            &       $27$            &       $25$            &       $21$                    \\
                &MCS                            &       $2^{15}$        &       $2^{15}$        &       $2^{16}$        &       $2^{17}$        &       $2^{18}$        &       $2^{19}$        &       $2^{19}$                \\      \hline
1.0     &$T_{in}$                               &       $0.6$   &       $0.6$   &       $0.7$    &      $0.7$   &       $0.7$   &       $0.8$   &       $0.8$           \\
                                &$N_T$                          &       $37$            &       $37$            &       $33$            &       $33$            &       $33$            &       $29$            &       $24$                    \\       -
                &MCS                            &       $2^{15}$        &       $2^{15}$        &       $2^{16}$        &       $2^{17}$        &       $2^{18}$        &       $2^{19}$        &       $2^{20}$                \\      \hline
1.75    &$T_{in}$                               &       $0.6$   &       $0.6$   &       $0.6$    &      $0.6$   &       $0.6$   &       $0.6$   &       $0.65$          \\
                                &$N_T$                          &       $37$            &       $37$            &       $33$            &       $33$            &       $33$            &       $20$            &       $22$                    \\       -
                &MCS                            &       $2^{15}$        &       $2^{15}$        &       $2^{16}$        &       $2^{17}$        &       $2^{18}$        &       $2^{20}$        &       $2^{20}$                \\      \hline
2.0     &$T_{in}$                               &       $0.01$  &       $0.01$  &       $0.025$  &      $0.025$ &       $0.3$   &       $0.4$   &       $0.5$           \\
                                &$N_T$                          &       $90$            &       $90$            &       $36$            &       $36$            &       $25$            &       $21$            &       $17$                    \\       -
                &MCS                            &       $2^{18}$        &       $2^{18}$        &       $2^{19}$        &       $2^{19}$        &       $2^{18}$        &       $2^{20}$        &       $2^{20}$                \\      \hline          \hline            \end{tabular}
\caption{Simulation parameters of the parallel tempering in
temperature: number of samples 2000, Monte Carlo Steps (MCS), number of
thermal bath $N_T$ spaced by $\Delta T=, 0.02$ or $0.025$.}
\label{tab:param}
\end{table}
\begin{table}[b!]
\begin{tabular}{| c | l || c | c | c | c | c |} \hline\hline
$T\downarrow $          &$L\rightarrow$                                 &       6               &       8               &       10               &      12              &       15              \\  \hline\hline
  0.2   &$D_{in}$       &       $1.99$  &       $1.999$ &       $2.00392$&      $1.981$ &       $1.981$ \\
                &$\Delta D_{in}$                &       $0.002$ &       $0.0006$        &       $0.00027$&      $0.003$ &       $0.003$ \\
                &$N_D$                          &       $21$            &       $21$            &       $37$             &      $37$            &       $37$            \\
                &MCS                            &       $2^{15}$        &       $2^{17}$        &       $2^{18}$         &      $2^{20}$        &       $2^{20}$        \\      \hline
0.3     &$D_{in}$                               &       $2.0034$        &       $2.026$ &       $2.0212$         &      $2.0256$        &       $2.028$ \\
                &$\Delta D_{in}$                &       $0.002$ &       $0.001$ &       $0.00037$&      $0.003$ &       $0.00025$\\
                &$N_D$                          &       $21$            &       $21$            &       $21$             &      $31$            &       $31$            \\
                &MCS                            &       $2^{15}$        &       $2^{17}$        &       $2^{17}$         &      $2^{17}$        &       $2^{18}$        \\      \hline
0.4     &$D_{in}$                               &       $2.05$  &       $2.06$  &       $2.057$  &      $2.06$  &       $2.062$ \\
                &$\Delta D_{in}$                &       $0.003$ &       $0.002$ &       $0.0007$         &      $0.00085$&      $0.0006$        \\
                &$N_D$                          &       $21$            &       $21$            &       $21$             &      $31$            &       $31$            \\
                &MCS                            &       $2^{15}$        &       $2^{17}$        &       $2^{17}$         &      $2^{17}$        &       $2^{18}$        \\      \hline
0.5     &$D_{in}$                               &       $2.06$  &       $2.06$  &       $2.06$   &      $2.026$ &       $2.026$ \\
                &$\Delta D_{in}$                &       $0.01$  &       $0.01$  &       $0.01$   &      $0.008$ &       $0.008$         \\
                &$N_D$                          &       $21$            &       $21$            &       $21$             &      $37$            &       $37$            \\
                &MCS                            &       $2^{15}$        &       $2^{17}$        &       $2^{17}$         &      $2^{18}$        &       $2^{18}$        \\      \hline \hline
\end{tabular}
\caption{Simulation parameters of the parallel tempering in $D$. Number of disordered samples: $1000$}
\label{tab:param2}
\end{table}
\\
\indent
Thermalization has been checked in three ways.
\label{fig:DISEGNO}
\begin{enumerate}
\item We have verified the symmetry with respect to zero of the site
overlap distribution function $P_J(q_s)$ for single random samples
(cf, Fig. \ref{fig:thermalization}). In absence of an external
magnetic field this must be symmetric for any choice of $\{J_{ij}\}$
realization.
\item
We have looked at the $t$-log behavior of the energy and we have
considered as thermalized those systems in which at least the last two
points coincide within the error,
cf. Fig. \ref{fig:thermalization}. This means that at least the last half of
the data in MCS can be used for computing statistical ensemble
averages.
\item
 we have checked the lack of variation on logarithmic time-windows of
all considered observables (e.g., $\xi_c$ and $\chi_{SG}$) on at least
two log points.
\end{enumerate}

\section{Numerical results.}
\subsection{Second order phase transition and universality}
\begin{figure}[t!]
\centering
\includegraphics[height=.95\columnwidth, angle=270]{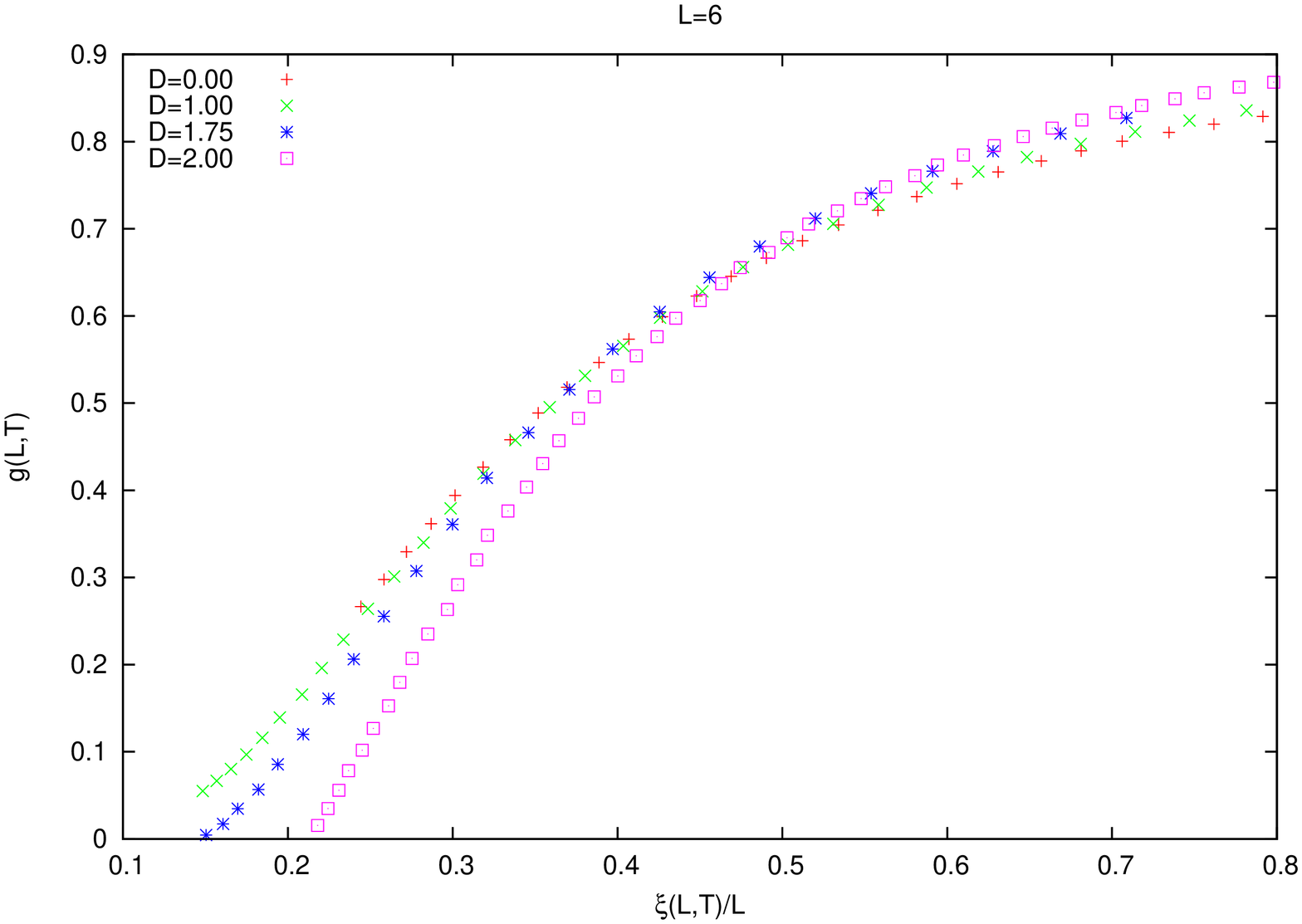}
\\
\includegraphics[height=.95\columnwidth, angle=270]{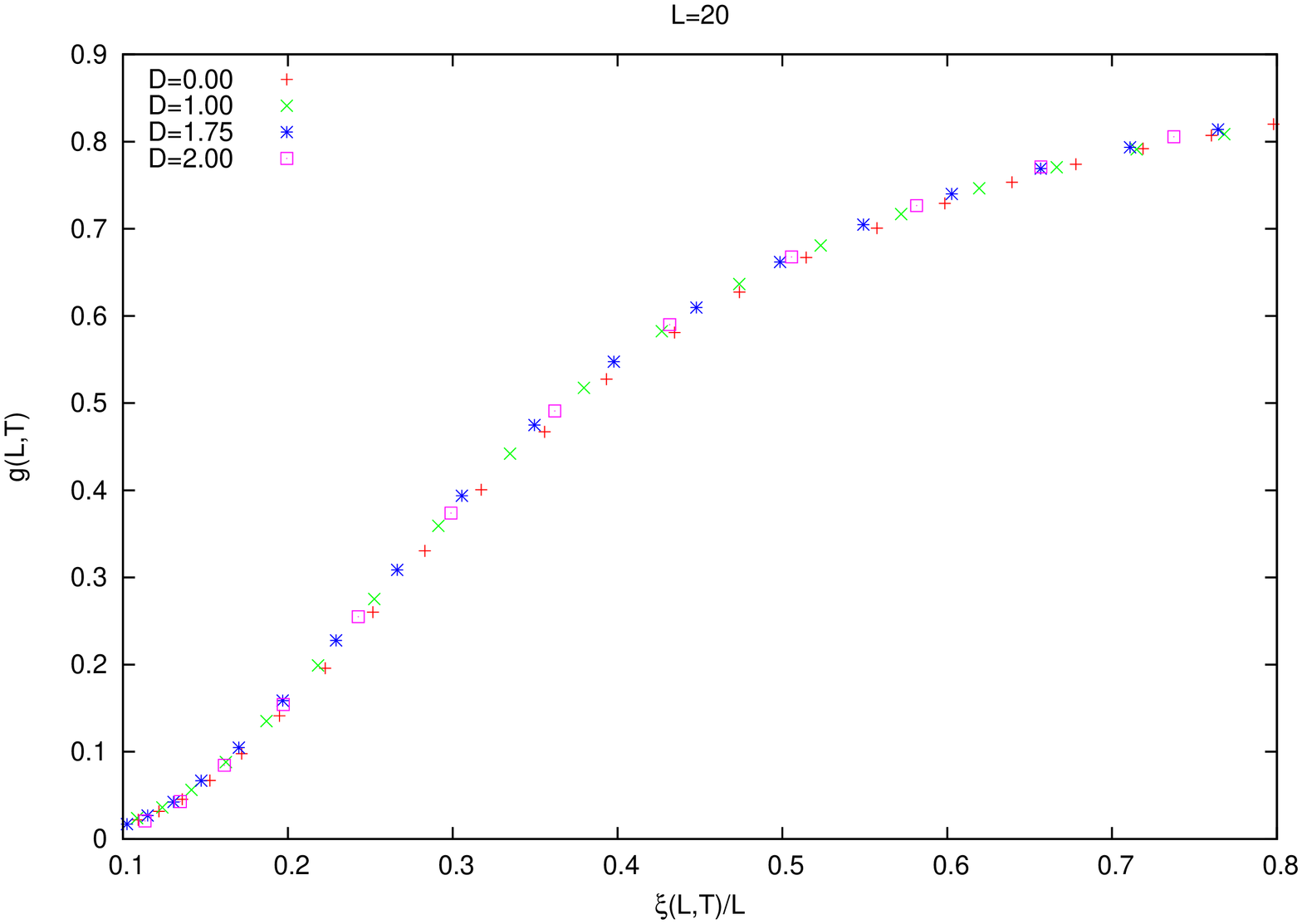}
\caption{Universal scaling function $g$ vs. $\xi_c/L$ at $L=6$ (left)
and $L=20$ (right) for all simulated $D$ values. At small size the
curves do not fall on top of each other as $D$ is too near the tricritical
point $D=1.75, 2$, whereas at large size their critical
behavior appears to be the universal for all  $D$.}
\label{fig:scale_Dall_L}
\end{figure}

\begin{figure}[t!]
\centering
\includegraphics[height=.95\columnwidth, angle=270]{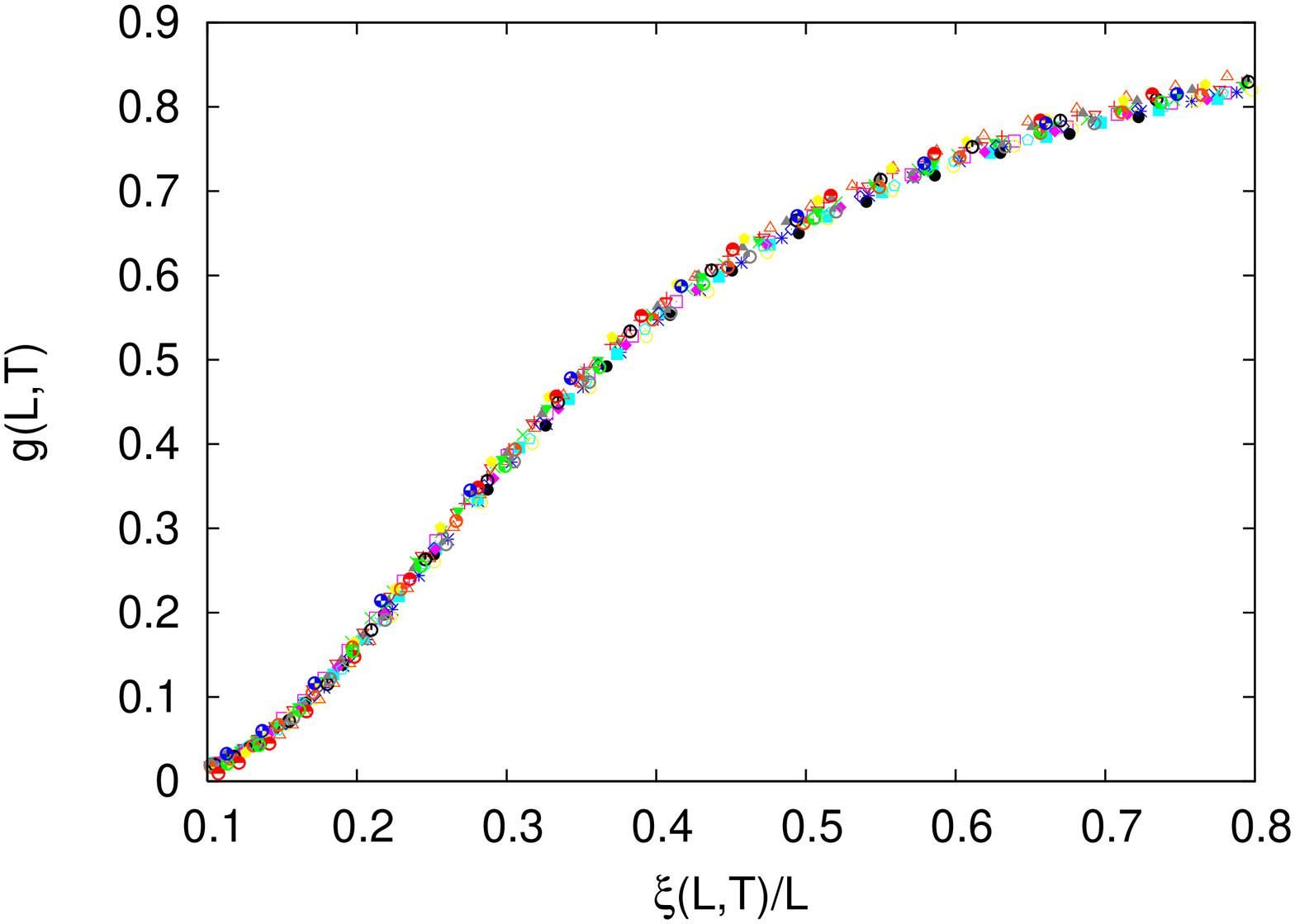}
\caption{Binder parameter $g(L,T_c(L))$ vs.
$\xi_c(L,T_c(L))/L$ in the critical region for different $D$ and
$L$. Values of $L=6,8,10,12$ for $D=2$ and $L=6$ for $D=1.75$ are
omitted.}
\label{fig:scale_Dall}
\end{figure}

\begin{figure}[t!]
\centering
\includegraphics[height=.95\columnwidth, angle=270]{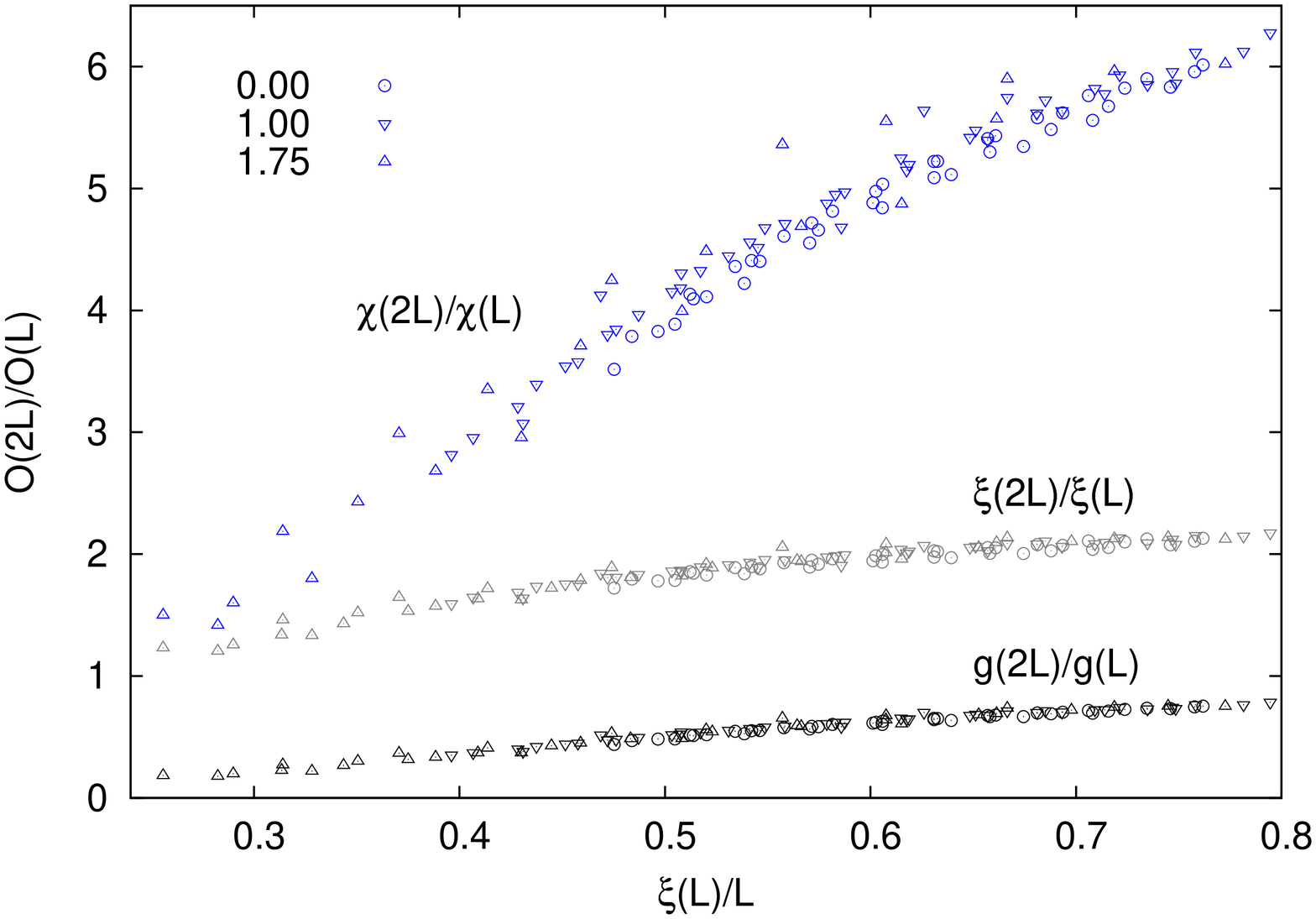}
\caption{ Scaling behavior of 
$F_{\chi_{\rm SG}}$, $F_{\xi}$ and $F_{g}$ (top to bottom)
vs. $\xi_c(L,T)/L$, cf., respectively, Eqs. (\ref{def:chisg}),
(\ref{f:xic}), (\ref{f:binder}) and (\ref{scaling}), at $D=0,1$,
$L/2L=6/12,8/16,10/20,12/24$ and at $D=1.75$, $L/2L=8/16,10/20,12/24$.}
\label{fig:Fchi}
\end{figure}

In Fig.  \ref{fig:xisuL} we present the $T$-behavior of $\xi_c/L$ for
different values of $D=0,1,1.75,2$ and $2.11$.  In the first four
cases the curves at different $L$ clearly cross, yielding evidence for a
non zero $T_c^L$.  From a FSS $T_c^L=T_c^\infty+a L^{-b}$ we can,
thus, extrapolate the critical temperature in the thermodynamic limit.
The $T_c^L$ crossing points between $L-2L$ curves are reported on
column $3$ of Tab. \ref{tab:qm}.  The FSS estimates are reported in
columns $2$ and $5$ of Tab. \ref{tab:crit}, where $L/L'$ couples are
chosen both as $L/2L$ (col. $2$) and as contiguous in the series
$L=6,8,10,12,16,20,24$ (col. $5$).  Analogous, though less precise
estimates, can be obtained by studying the behavior of the Binder
cumulant $g$, cf. Eq. (\ref{f:binder}). 
 Applying both the quotient
and the conventional FSS methods we can, eventually, obtain two estimates
for the critical exponents.
\begin{table}[!b]
\begin{tabular}{|| c | c | c | c | c | c | c ||} \hline
$D$         & $L-L^\prime$    &    $T_c(s)$    &    $Q_{\partial_\beta \xi}(s)$&$\nu(s)$         &        $Q_{\chi}(s)$      &    $\eta(s)$    \\ \hline\hline
                &   $6-12$            &$1.02(4)$    &    $1.35(1)$&$2.31(1)$             &        $5.1(1)$        &    $-0.34(4)$            \\
$0.0$     & $8-16$            &$0.99(6)$        &    $1.31(2)$&$2.58(5)$            &        $5.1(1)$         &    $-0.36(3)$            \\   
               & $10-20$        &$1.0(1)$        &    $1.35(4)$&$2.3(1)$                &        $5.2(1)$        &    $-0.39(4)$            \\   
               & $12-24$        &$0.98(9)$    &    $1.33(2)$&$2.43(7)$            &        $5.1(1)$        &    $-0.35(4)$        \\\hline
        &$\infty$            &    $1.01(1)$             &         {\phantom{$1.34(1)$}}    &$2.34(3)$            &      {\phantom{$5.13(1)$}}        &    $-0.36(1)$                         \\\hline\hline   
$D$         & $L-L^\prime$    &    $T_c(s)$    &$Q_{\partial_\beta \xi}(s)$        &    $\nu(s)$         &         $Q_{\chi}(s)$     &    $\eta(s)$        \\\hline
        & $6-12$        &$0.894(9)$    &$1.32(1)$                    &    $2.51(4)$        &        $4.9(3)$         &    $-0.29(9)$        \\
$1.0 $    & $8-16$        &$0.895(9)$    &$1.396(6)$                &    $2.08(1)$        &        $4.8(4)$         &     $-0.26(1)$            \\   
        & $10-20$        &$0.877(9)$    &$1.271(7)$                &    $2.89(2)$        &        $5.1(5)$         &    $-0.33(1)$            \\
        & $12-24$        &$0.86(1)$    &$1.35(1)$                    &    $2.29(2)$        &        $5.1(5)$         &    $-0.3(1)$                \\\hline
        & $\infty$&    $0.88(1)$    &&$2.45(1)$    &                    &    $-0.31(2)$    \\\hline\hline   
$D$         & $L-L^\prime$&$T_c(s)$&$Q_{\partial_\beta \xi}(s)$    &$\nu(s)$&$Q_{\chi}(s)$     &    $\eta(s)$    \\\hline
            &$6-12$    &$0.715(7)$& $1.41(1)$    &{\phantom{$1.95(3)$}}*&    $4.7(5)$ &    {\phantom{ $-0.2(1)$}}*            \\
$1.75$     &$8-16$    &$0.679(9)$&$1.37(1)$    &$2.12(4)$        &$4.8(5)$ &     $-0.3(1)$            \\   
               &$10-20$    &$0.67(1)$&$1.34(3)$    &$2.4(1)$ &    $5.0(6)$         &    $-0.3(1)$        \\
               &$12-24$ &$0.68(1)$&$1.38(1)$    &$2.11(3)$&    $4.9(5)$         &    $-0.3(1)$        \\\hline
&$\infty$    &    $0.69(1)$    &     &$2.20(3)$&     &    $-0.30(1)$            \\\hline\hline   
$D$  & $L-L^\prime$&$T_c(s)$    &    $Q_{\partial_\beta \xi}(s)$&$\nu(s)$&$Q_{\chi}(s)$&$\eta(s)$    \\\hline
     & $6-12$    &$0.593(7)$    &    $1.59(4)$    &{\phantom{$1.49(3)$}}    *    &$9.5(9)$&{\phantom{$-1.2(1)$}}*    \\
$2.0$& $8-16$    &$0.569(8)$    &    $1.47(3)$    &{\phantom{$1.81(6)$}}    *    &$18(2)$ &{\phantom{$-2.2(2)$}}*    \\   
     & $10-20$    &$0.54(1)$    &    $1.37(3)$    &{\phantom{$2.12(6)$}}    *    &$16(2)$ &{\phantom{$-2.0(2)$}}*\\
     & $12-24$    &$0.54(1)$    &    $1.34(4)$    &{\phantom{$2.3(1)$}}    *    &$10(2)$ &{\phantom{$-1.3(3)$}}*\\\hline
  & $\infty$ &    $0.56(1)$    &                    &                               &        &             \\\hline\hline   
\end{tabular}
\caption{Critical temperature and exponents are calculated with QM:
for $D=0.00$, $D=1.00$ and $D=1.75$, through a FSS analysis of the
values of $Q_{\partial_\beta \xi}(s)$ and $Q_{\chi}(s)$ for
$s=L'/L=2$. Cells with * mean that quotients are
computed on sizes too small to significantly represent the asymptotic
behavior with $L$.}
\label{tab:qm}
\end{table}
\\
\indent
Before applying these methods, though, we must check if we can exclude
cross-over effects as the chemical potential $D$
is varied due to FS.
Since we are in presence of a tricritical point, signaled,
among others, by the weird behavior of $\xi_c/L$ at $D=2.11$,
cf. Fig. \ref{fig:xisuL}, we should control how it influences the
results as it is approached along the continuous transition line
increasing $D$. 
 In the mean-field approximation, indeed, at the
tricritical point the coefficient of the fourth order term in the SG
free energy action goes to zero and the sixth order term becomes
relevant for the critical behavior, as shown in Ref.
[\onlinecite{CLPRL02}]. This is a typical behavior of Blume-Emery-Griffiths-Capel-like systems
\cite{Riedel72,ZZ} that might hinder the determination of the critical
behavior in the neighborhood of the tricritical point for sizes that
are ``not large enough''.
\begin{widetext}
\begin{minipage}{0.98\textwidth}
\centering
\includegraphics[height=0.95\textwidth, angle=270]{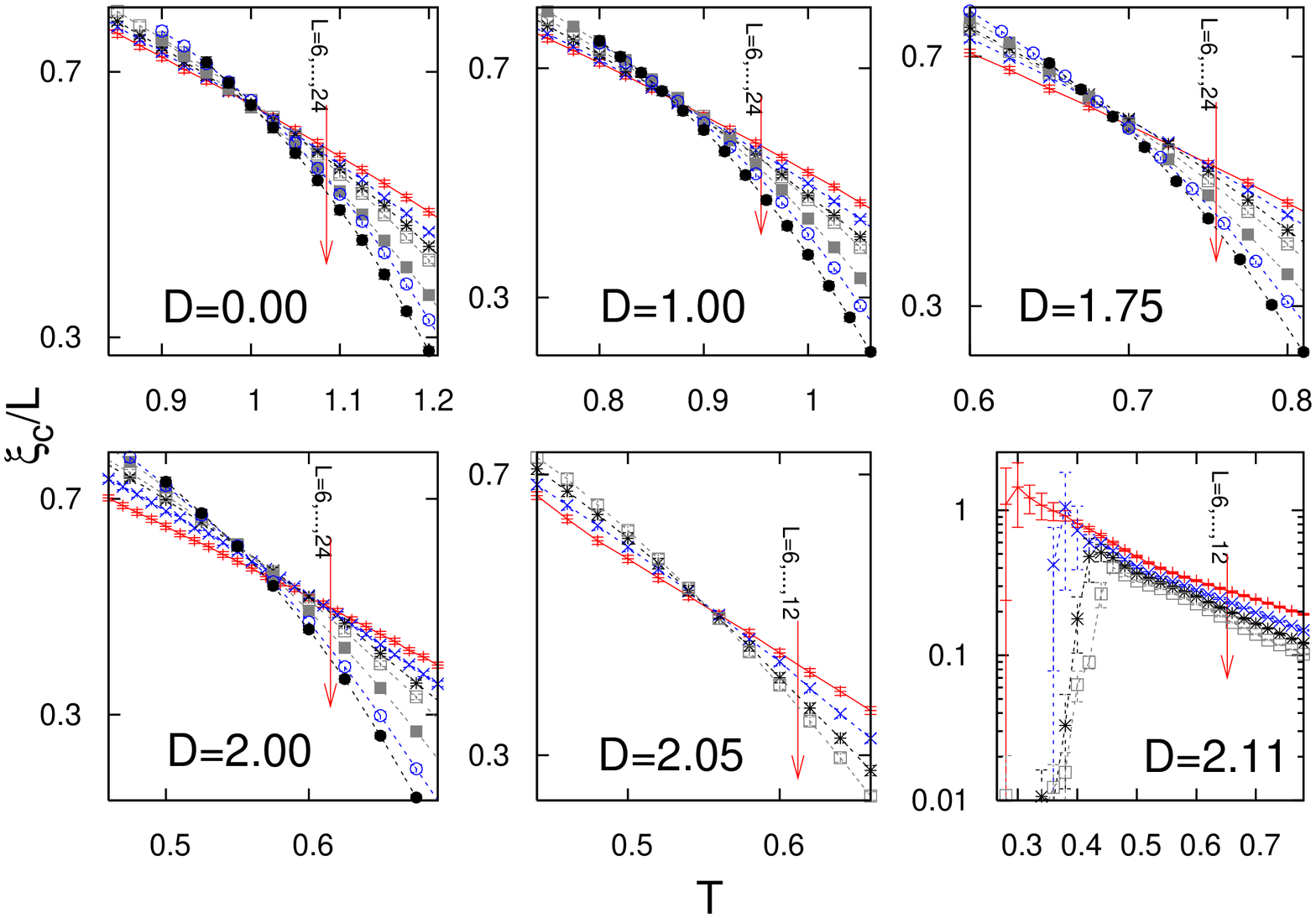}
\figcaption{Scaling functions $\xi_c/L$ vs. $T$ for different values
$D$. For $D=0,1,1.75,2$ ($L=6,8,10,12,16,20,24$)  evidence for a
continuous phase transition is found in the region of scale
invariance. At $D=2.11$ ($L=6,8,10,12$) no crossing is observed and,
at low $T$, $\xi_c/L\to 0$.}
\label{fig:xisuL}
\end{minipage}
\end{widetext}
 To estimate and control FS effects we use the scaling
methods introduced in Sec. \ref{sec:uniFSS} and compare different
universal FSS functions.  In Fig. \ref{fig:scale_Dall_L} we plot the
Binder parameter $g$ vs. the rescaled correlation length $\xi_c/L$ at
all simulated values of the chemical potential $D$ both for a small
($L=6$, top) and a large ($L=20$, bottom) system. 
\noindent In the top plot one
can easily observe that as the tricritical value of $D$ is approached
($2.05<D_{3c}<2.11$) for $L=6$ the curves do not overlap with each
other signaling an apparent lack of universality. At large enough
sizes, instead, all curves are superimposed (bottom plot of
Fig. \ref{fig:scale_Dall_L}, $L=20$), demonstrating that universality
holds along the whole continuous transition line and that, because of
strong FS effects, a crossover occurs and the analysis limited to (or
including also) too small sizes can hinder the prediction of the
asymptotic behavior.
\begin{table}[!b]
\begin{tabular}{|| c | c | c | c || c | c | c ||}\hline
$D$         &    $T_c$&$\nu$&$\eta$& $T_c$&    $\nu$&    $\eta$    \\\hline\hline
$0.00$    &    $1.01(1)$&$2.34(3)$&$-0.36(1)$&   $1.0(1)$  & $2.5(2)$&$-0.37(2)$        \\
$1.00$    &    $0.88(1)$&$2.45(1)$&$-0.31(2)$&   $0.8(1)$   &$2.6(5)$    &$-0.31(2)$        \\
$1.75$     &    $0.68(2)$&$2.20(3)$*&$-0.30(1)$*&$0.6(1)$   &$2.6(6)$&$-0.30(4)$    \\
$2.00$    &    $0.56(1)$&$\dagger$&$\dagger$&      $0.5(1)$ &    $2.3(2)$    &$-0.31(2)$        \\ \hline\hline
\end{tabular}
\caption{Critical temperature and exponents calculated via QM $Q_{\p
\beta \xi}(s,T_c(s))$ and $Q_{\chi_{SG}}(s,T_c(s))$ (cols. $2$,$3$ and
$4$) and via standard FSS analysis of the behavior of $\log\p
\xi_{c}(L,T_c(L))/\p \beta$ and $\log\chi_{SG}(L,T_c(L))$ (cols. $5$,
$6$ and $7$). *: estimated through QM without $L=6$. $\dagger$: not
estimated by QM.}
\label{tab:crit}
\end{table}
\begin{figure}[t!]
\centering
\includegraphics[width=.95\columnwidth]{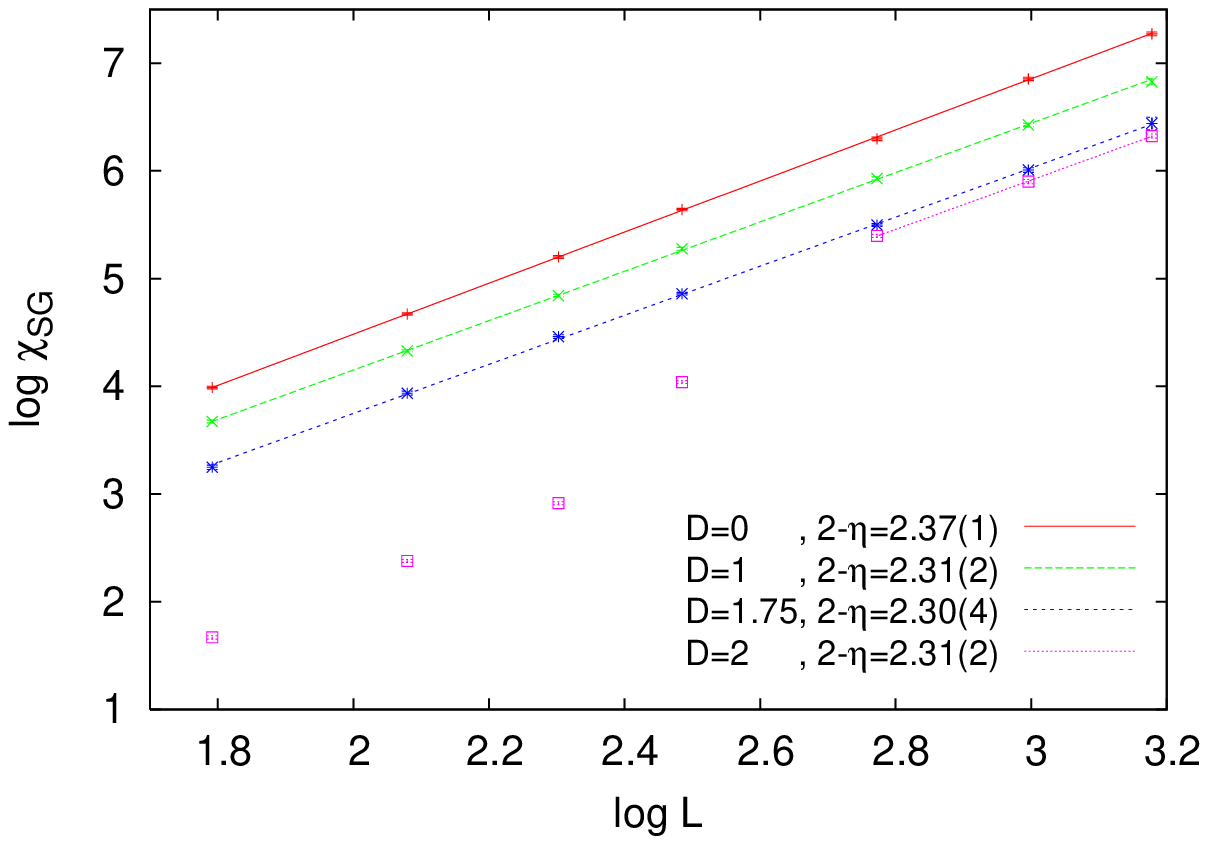}
\caption{$\chi_{\rm SG}$ vs. $L$ in log-log plot for
$D=0,1,1.75,2$. For $D=2.00$, near the tricritical point, a cross-over
is evident from small ($L\lesssim 12$) to large ($L>12$) sizes.  The
quotient method does not yield reliable estimates because of a
crossover in the scaling functions in the range of probed sizes.}
\label{fig:chilog_2}
\end{figure}
\\
\indent
The same effect is clearly shown in Fig. \ref{fig:chilog_2} where the
size dependence of spin-glass susceptibility at criticality is shown.
As $D$ increases towards $D_{3c}$ there appears to be a crossover
in the scaling moving from small to large sizes and induces wrong
asymptotic values of the critical indices.
 We, thus, did not make use of the small values of $L$ for
$D\simeq D_{3c}$, namely $L=6$ at $D=1.75$ and $D=6,8,10$ and $12$ at
$D=2$, to interpolate the values of the critical exponents, as they
induce a wrong estimate as the limit $L\to \infty$ is performed.
\\
\indent
As a
test for the eye, in Fig. \ref{fig:scale_Dall}, we display $g$
vs. $\xi_c/L$ for all $D$ and $L$ values employed for our FSS
analysis. Without the smallest sizes near the tricritical point,
universality appears quite tidy.  In Fig. \ref{fig:Fchi}
we parametrically plot the universal FSS functions $F_\xi$,
$F_{\chi_{\rm SG}}$ and $F_g$, cf. Sec. \ref{sec:uniFSS}, vs. $\xi_c/L$,
as well, for the same simulated systems.
\\
\indent
The critical values of the temperature and the exponents $\eta$ and
$\nu$ are shown in Tab. \ref{tab:crit} both for the QM and the
canonical FSS methods. Due to the FS cross-over no interpolation was
possible with QM at $D=2$. We, thus, provide only one estimate for the
indeces.
\begin{table}[t!]
\begin{tabular}{|| l | c | c ||}
\hline\hline
Model           &       $\nu$   &       $\eta$          \\ \hline\hline
SG 3D bd \cite{Jorg06}  &       2.22(15)        &       -0.349(18)      \\ \hline
SG 3D bd \cite{HPV08}&  2.53(8) &       -0.384(9)               \\ \hline
EA 3D \cite{BCPRB00}&   2.15(15)        &       -0.337(15)      \\ \hline
EA 3D \cite{MPRPRB98}   &2.00(15)       &       -0.36(6)                \\ \hline
\hline
\end{tabular}
\caption{Critical Indices of EA models in literature}
\label{tab:EA}
\end{table}
\\
\indent
As one can see, comparing with estimates of critical exponents
summarized in Tab. \ref{tab:EA}, the system appears to be in the same
universality class of the Edwards-Anderson model (corresponding to the
$D=-\infty$ limit of our model).\cite{BCPRB00,janus1, HTPV1,JK1}
\\ \indent In Fig. \ref{fig:xisuL} we also plot $\xi/L$ at $D=2.05$
and $D= 2.11$ for $L=6,8,10,12$. In the first case we obtain a
$T_c=0.553(7)$, though no analysis of the critical exponents can be
performed because of FS effects. In the latter case no evidence is
found for a second order phase transition,
cf. Fig. \ref{fig:xisuL}. As $T\lesssim 0.5$ is approached, moreover,
$\xi$ even appears to scale weaker than $L$.  We will see in the
following why this comes about.
%

%%%%%%%%%%%%%%%%%%%%%%%%%%%%%%%%%%%%%%%%%%%%%%%%%%%%%%%%%%%%%%%%%%%%

\subsection{First order phase transition}
\begin{figure}[t!]
\centering
\includegraphics[width=.95\columnwidth]{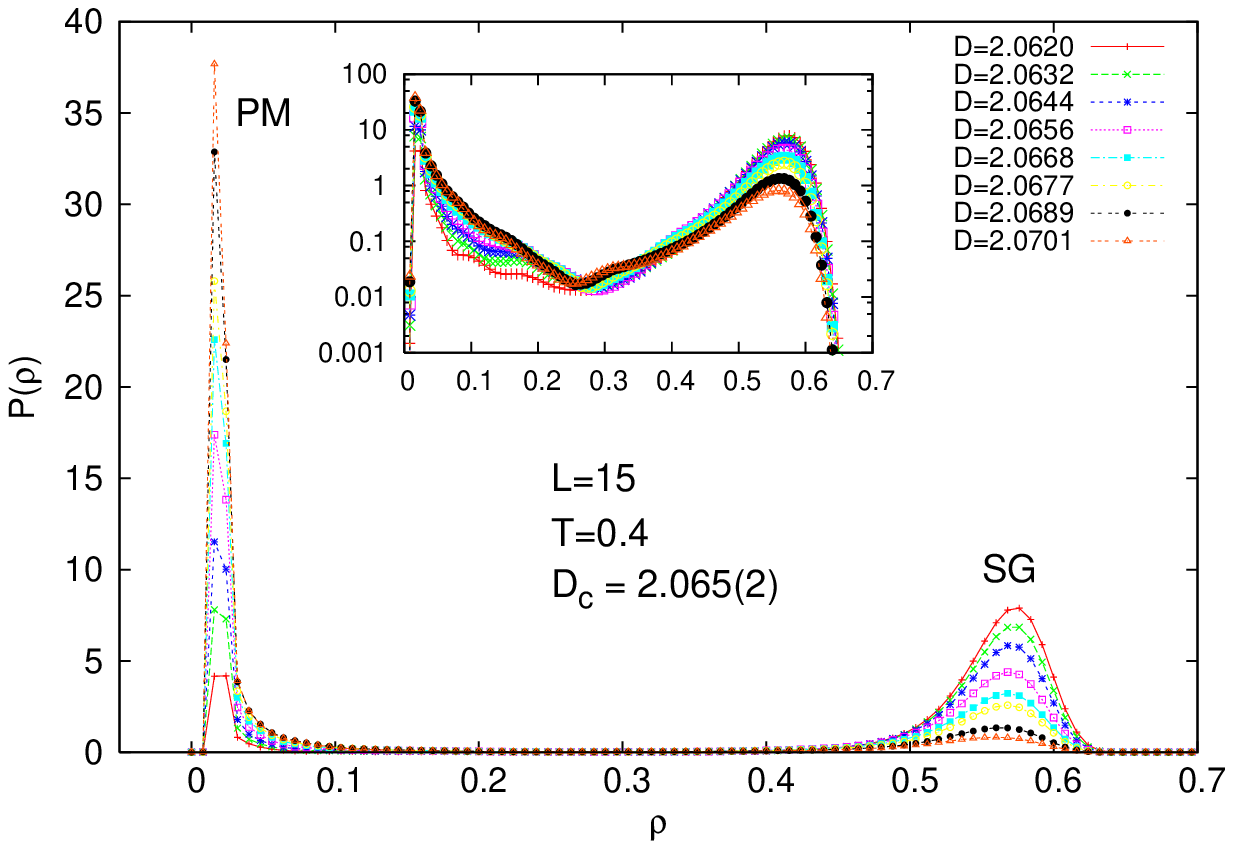}
\caption{Density distribution $P(\rho)$, $L=15$, across the
coexistence region at $T=0.4$: two peaks develop at $\rho_{PM}$ and
$\rho_{SG}$. As $D$ increases the thermodynamically relevant phase
(lowest free energy) passes from SG to PM in a first order phase
transition. The dominant phase corresponds to the one with larger
probability, i.e., larger integral of the peak. As the peak at
$\rho_{SG}$ vanishes the system is in a purely PM phase.  Inset:
$P_{15}(\rho)$ on $y$-Log scale.}
\label{fig:PrhoT04}
\end{figure}

\begin{figure}[t!]
\centering
\includegraphics[height=.95\columnwidth, angle=270]{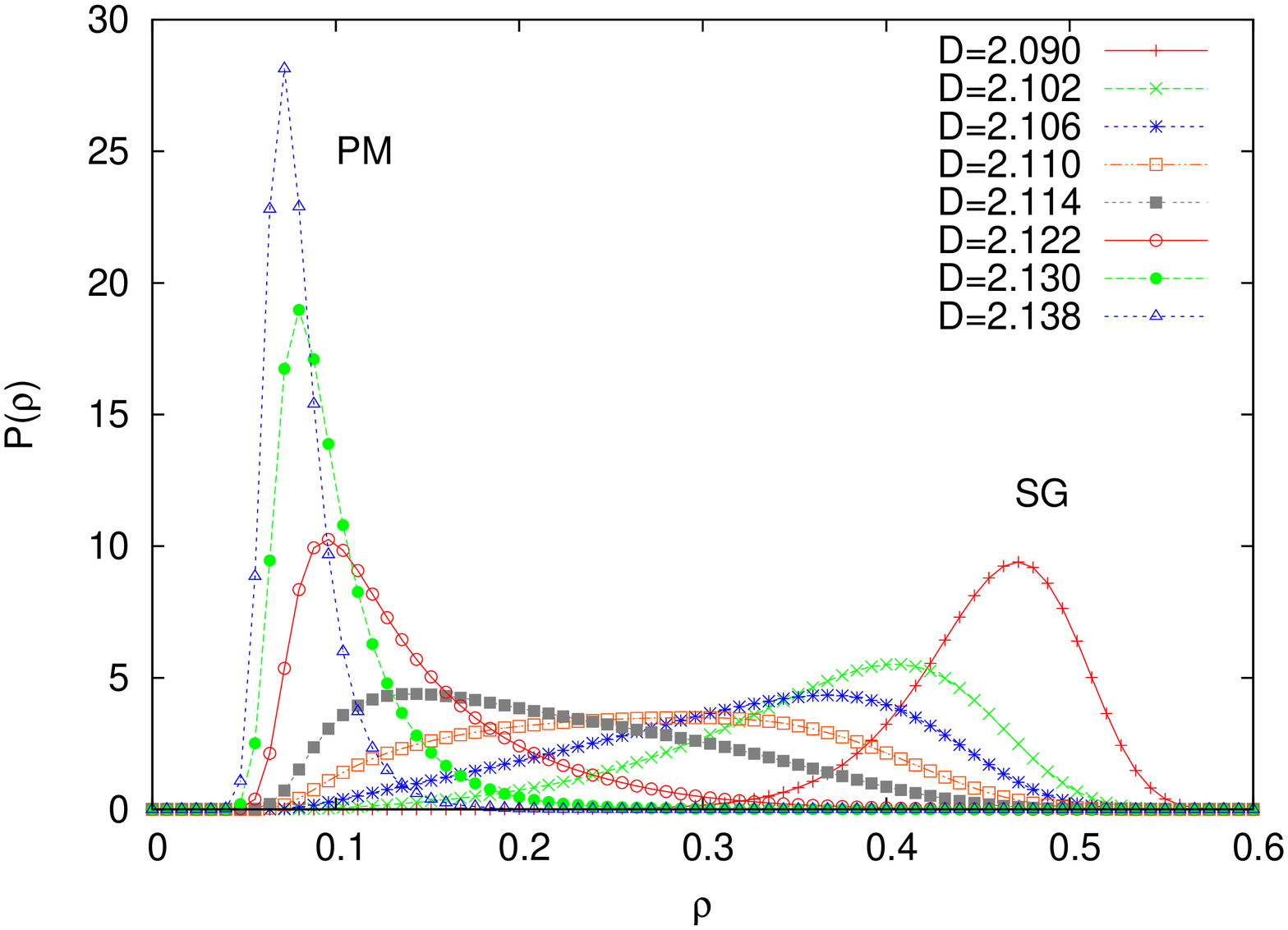}
\caption{$P(\rho)$ in the coexistence region at $T=0.5$ and
$L=15$. The two peaks are hard to distinguish and the coexistence
region is rather narrow.}
\label{fig:PrhoT05}
\end{figure}

\begin{figure}[t!]
\centering
\includegraphics[height=.95\columnwidth, angle=270]{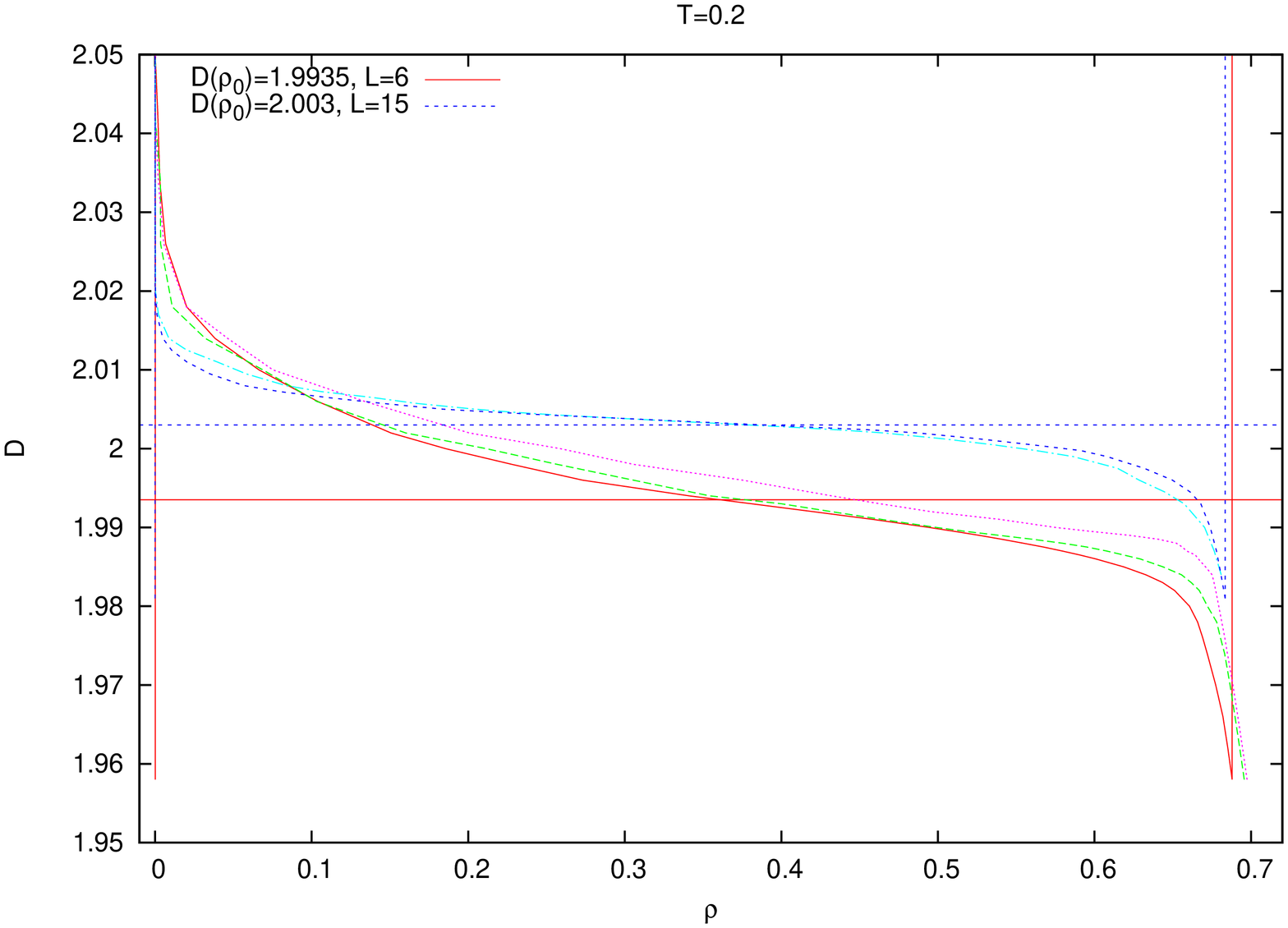}
\caption{Maxwell construction at $T=0.2$ in the $\rho$, $D$
plane. $L=6,8,10,12,15$ from bottom to top as $\rho\gtrsim 0.6$ 
(right). The almost vertical lines at the small and large density
sides are the interpolated pure phase (PM left, SG right)
behaviors. As an instance the critical $D$ values for the equal
distance construction are plotted at $L=6$ (lower horizontal line) and
at $L=15$ (higher horizontal line).}
\label{fig:maxwell}
\end{figure}

Across a second order transition the system undergoes a transformation
from a PM pure phase to a SG pure phase.  As far as  the density
distribution $P(\rho)$ is concerned, a pure phase corresponds to a
single-peaked distribution. As two peaks appear, the system exists
both in PM (low $\rho$) and SG (high $\rho$) coexisting phases and we
are in the neighborhood of a first order phase transition.  In FS
systems the peaks are not delta-shaped but become sharper and sharper
as $L$ increases.  At finite $L$, thus, $P(\rho)$ is a good order
parameter that drives the first order kind of transition: varying
$D,T$, the system undergoes a transition with a discontinuous jump in
$\rho$  and the ``thermodynamic'' average values of
$\rho$ are obtained by looking at the peaks of its distribution.
\\
\indent
 In Fig. \ref{fig:PrhoT04} we show the behavior of the density
distribution through the first order transition at $T=0.4$.  The FS
first order transition points can be determined with the four methods
mentioned in Sec. \ref{sec:fo}, as we will show below.  The spinodal
lines at given $L$ are estimated by looking at the $D$ values at which
a secondary peak arises.
Since the region of phase coexistence corresponds to an inverse freezing
transition, we performed PT
 simulations at finite $T$, changing $D$.
Indeed, in our model, we will see that the first order transition line
displays a reentrance \cite{CLPRL05,LPM} due to the existence of a
"fluid'' (PM) phase with an entropy lower than the one of the glassy
phase.
\\
\indent
For what concerns the estimate of $D_c(T)$ the method of {\em equal
weight} introduced in Sec. \ref{sec:fo}, cf. Eq. (\ref{f:fo_ew}) works
quite well for data collected at $T\leq 0.4$, because the two peaks
are very well separated as soon as they appear,
cf. Fig. \ref{fig:PrhoT04}, and the estimate is robust against
reasonable changes of $\rho_0$ (see inset of Fig. \ref{fig:PrhoT04}).
As $T$ increases towards the tricritical value, however, the PM and SG
values of the density approach each other.  At $T=0.5$,
cf. Fig. \ref{fig:PrhoT05}, we thus have the problem that the
distributions of the densities of the two phases are overlapping also
for the largest simulated size.  In that case, seen the arbitrariness
of choosing $\rho_0$, we actually determine the transition point as
the $D$ value at which the peaks have the same height. This is a  rough estimate
but yields no difference w.r.t., e.g., fitting the two peaks separately
and computing the areas under the interpolating curves.  In
Tab. \ref{tab:Dc_ew} we report for all simulated sizes and
temperatures the estimated values of the critical points obtained by
this method, together with the spinodal points.
\\ \indent These results can be cross-checked using the methods based
on the Maxwell construction, cf. Sec. \ref{sec:fo} and
Fig. \ref{fig:maxwell}. The pure phase behaviors $D_{PM,SG}(\rho)$ are
interpolated in the coexistence region by a polynomial fit on those
points for which no double peak is present in the $P(\rho)$.  At any
given $L$ we look at the value $D=D_c$ such that (equal
distance)
 \BEQ \rho_{SG}{D_c}-\rho(D_c) = \rho(D_c) - \rho_{PM}(D_c) \nn \EEQ 
and at
the value of $D=D_c$ at which the areas between $D_c$ and $D(\rho)$ to
the left and to the right of their crossing point are equal, i.e., 
\BEQ \Delta\mathcal{A}(D_c)=\int_0^{\rho(D_c)}\!\!\!\!\!\!
d\rho'\!(D(\rho')-D_c)-\int_{\rho(D_c)}^1\!\!\!\!\! \!d\rho'\!(D(\rho')-D_c).
\EEQ
 is zero.
\begin{table}[t!]
\begin{tabular}{|| c | c | c | c ||}\hline\hline
$T$ & $D_c $ & $D_{SP}^{PM}$ & $D_{SP}^{SG}$\\
\hline
 &${\phantom .} $&\vspace*{-.4 cm} & \\
\hline
 $0.2$    &    $2.0031(1)$     &       $1.9833(2)$     &       $2.024(1)$ \\
 $0.3$   &     $2.032(3)$              &       $2.015(1)$              &       $2.043(5)$      \\
 $0.4$   &     $2.060(1)$              &       $2.046(2)$              &       $2.092(5)$      \\
 $0.5$   &     $2.106(1)$              &       $2.097(4)$              &       $2.143(4)$      \\ \hline\hline
\end{tabular}
\caption{Results of the first order phase transition: a fine tuning of
the parameters $\{D_i\}$ is needed in order to establish the critical
values $D_c$, $D_{SP}$ and $D_{SG}$.}
\label{tab:Dc_ew}
\end{table}
\begin{figure}[t!]
\centering
\includegraphics[height=.95\columnwidth, angle=270]{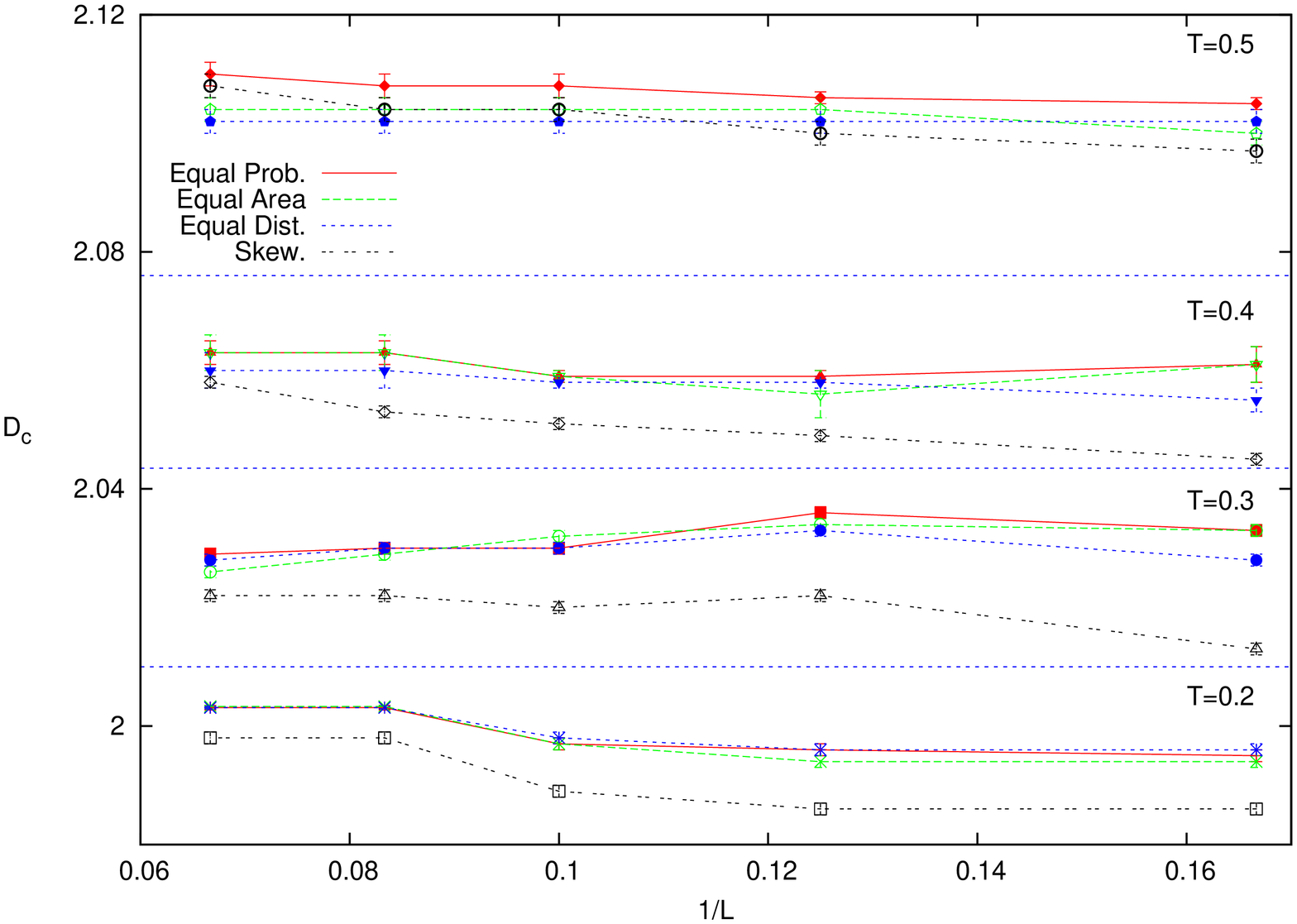}
\caption{Estimates of $D_c$ vs. $1/L$ for $L=6,8,10,12,15$ at $T=0.2,
0.3, 0.4,0.5$ (bottom up) obtained by the four methods described in
Sec. \ref{sec:fo}.}
\label{fig:Dc4}
\end{figure}
\\
\indent We, eventually, compute the skewness of double peaked $P(\rho)$ as $D$
changes, looking at the $D=D_c$ point for which $\zeta(\rho(D_c))=0$.
Since the two peaks of $P(\rho)$ at finite size appear to be of
different shape (SG broader, PM narrower),
cf. Figs. \ref{fig:PrhoT04}, \ref{fig:PrhoT05}, the point at which the
skewness is zero appears to be slightly different from the $D_c$
values computed with the previous three methods.  In
Fig. \ref{fig:Dc4} we plot at different temperatures the FS values
of $D_c(L)$ with the four methods.  The equal weight methid and the
two Maxwell construction methods yield consistent results.  For
$T=0.4, 0.5$ the estimate of $D_c$ by the symmetric distribution
method displays a growing behavior in $1/T$ that does not allow for a
consistent $L\to\infty$, cf. Fig. \ref{fig:maxwell} first and second
panel from top, whereas at lower temperature, where the interpolated
thermodynamic limit is stable the value is smaller than the other
estimates.
\\
\indent 
Summarizing, in Tab. \ref{tab:Dc3} we report the estimates of the first order
critical point obtained by means of the four methods.  
\begin{table}[!t]
\begin{tabular}{| l | c | c | c | c | }\hline \hline
T      &    $D_c[P(\rho)]$  &    $D_c[\rho_{\rm ed}]$     &    $D_c[\Delta\mathcal{A}=0)]$    &    $D_c[\zeta=0]$ \\ \hline
0.2    &    2.0031(1)   &    2.0033(2)         &     2.0031(2)                               &    1.991(2)\\
0.3       &      2.032(3)           &    2.031(2)       &      2.030(1)                         &    2.020(2)\\
0.4       &      2.060(1)           &    2.060(1)           &       2.058(1)                         &   x\\
0.5       &      2.106(1)           &    2.103(3)          &      2.102(1)                         &    x\\ \hline
\end{tabular}
\caption{Evaluation of the first order critical point with the method
of equal weight (col. 2), equal distance (col. 3), equal area (col. 3)
and zero skewness.}
\label{tab:Dc3}
\end{table}

\subsection{Phase diagrams and inverse freezing}
\begin{figure}[t!]
\centering
\includegraphics[width=.95\columnwidth]{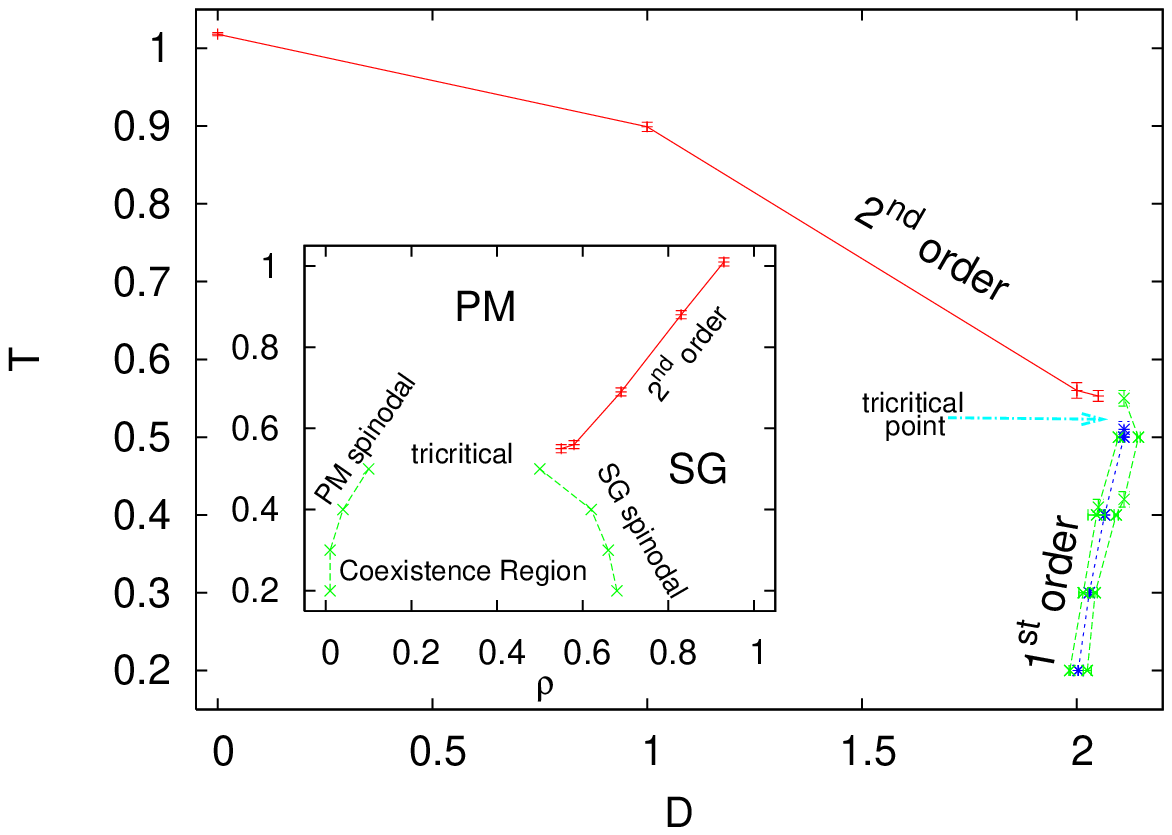}
\caption{Phase diagram in $D,T$: second order transition and an
inverted first order phase transition occur. In the latter case also the spinodal lines are
reported (dashed). Inset: $T,\rho$ phase diagram.}
\label{fig:phdi}
\end{figure}

\begin{figure}[t!]
\centering
\includegraphics[width=.95\columnwidth]{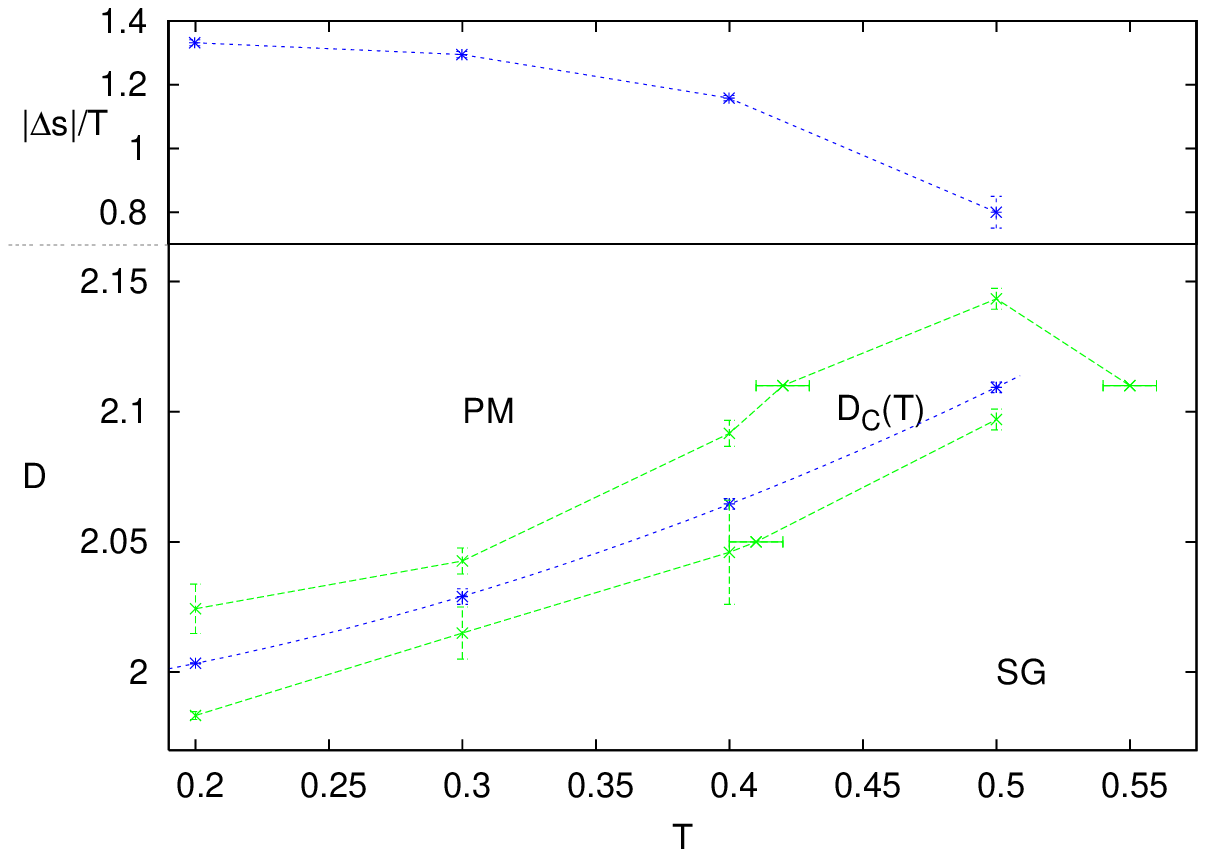}
\caption{Detail of inverse freezing region, interpolation of transition line
$D_c(\infty,T)$ (dotted), spinodal lines (dashed). The error bars are
the FSS of the minimal interval in $T$ and $D$ at each $L$ needed to
identify the crossings in $\xi_c/L$ curves (for continuous
transitions) or compare the areas under $P_{N}(\rho)$ for first order transition. In the top 
inset we show the  latent heat $|\Delta s|/T$ along the first order transition
line. }
\label{fig:phdi_det}
\end{figure}

Phase diagrams are plotted in Fig. \ref{fig:phdi}.
  In the $D,T$ plane we observe a pure SG phase at
low $T$ and $D\lesssim 2$. Increasing the temperature the continuous
transition to the pure PM phase is denoted by a full line connecting
the five numerical estimates of $T_c$ obtained by simulations at
$D=0,1,1.75,2$ and $D=2.05$.  We found no evidence for a continuous
phase transition at $D=2.11$.  Beyond $(D,T)=(2.05,0.53(2))$ a
tricritical point is placed.  Beside changing to a first order
transition, for lower $T$ also a reentrance in the $T_c(D)$ line
occurs.  The warmest first order point for which we have an estimate
is $(D,T)=(2.109(2),0.5)$.  In Fig.  \ref{fig:phdi_det} a
detail of the phase coexistence region is plotted (inside the
grey-dotted lines).  In the inset of Fig. \ref{fig:phdi} we plot the $(\rho,T)$
diagram. Below $T=0.53(2)$ no pure phase exists with an average $\rho$
in between the dashed-grey curves.
\indent
The inverse freezing takes place between a SG of high density to an almost empty PM
(e.g., at $T=0.4$, in the coexistence region $D\in[2.046(2):2.092(5)]$,
$\rho_{SG}\simeq 0.52$ and $\rho_{PM}\simeq 0.03$). The few active
sites do not interact with each other but only with inactive neighbors
and this induces zero magnetization and overlap. The corresponding PM
phase at high $T$ has, instead, higher density (e.g.,
$\rho_{PM}(D=2,T=0.6)=0.4157(2)$, $\rho_{PM}(D=2.11,T=0.6)=0.596(2)$)
and the paramagnetic behavior is brought about by the lack of both
magnetic order (zero magnetization) and blocked spin configurations
(zero overlap).
\\
\indent
Using Eq. (\ref{eq:CC}), from the
knowledge of $\Delta \rho$ and the numerical estimate of $dD/dT$ we
are able to evaluate the latent heat employed in the transition, that
we plot as a function of temperature in the top inset of
Fig. \ref{fig:phdi_det}.

\section{Nature of the SG phase.}
\label{sec:theories}
\begin{figure}[!t]
\centering
\includegraphics[height=.97\columnwidth, angle=270]{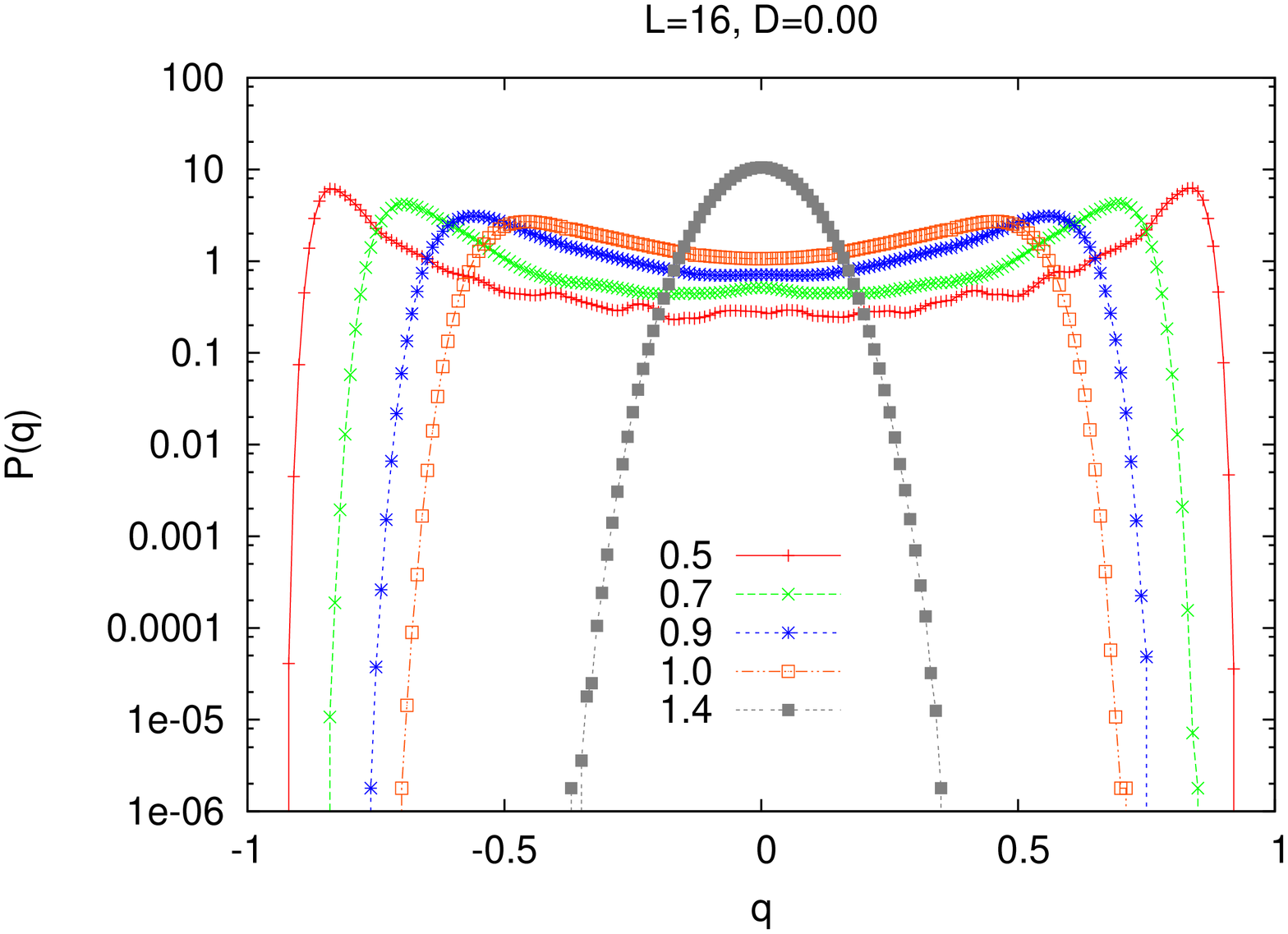}
\caption{Behavior of the overlap distribution $P(q)$
through the second order phase transition and in
the low-temperature phase for $L=16$. }
\label{f:pqs_L16_D0}
\end{figure}

\begin{figure}[!t]
\centering
\includegraphics[height=.95\columnwidth, angle=270]{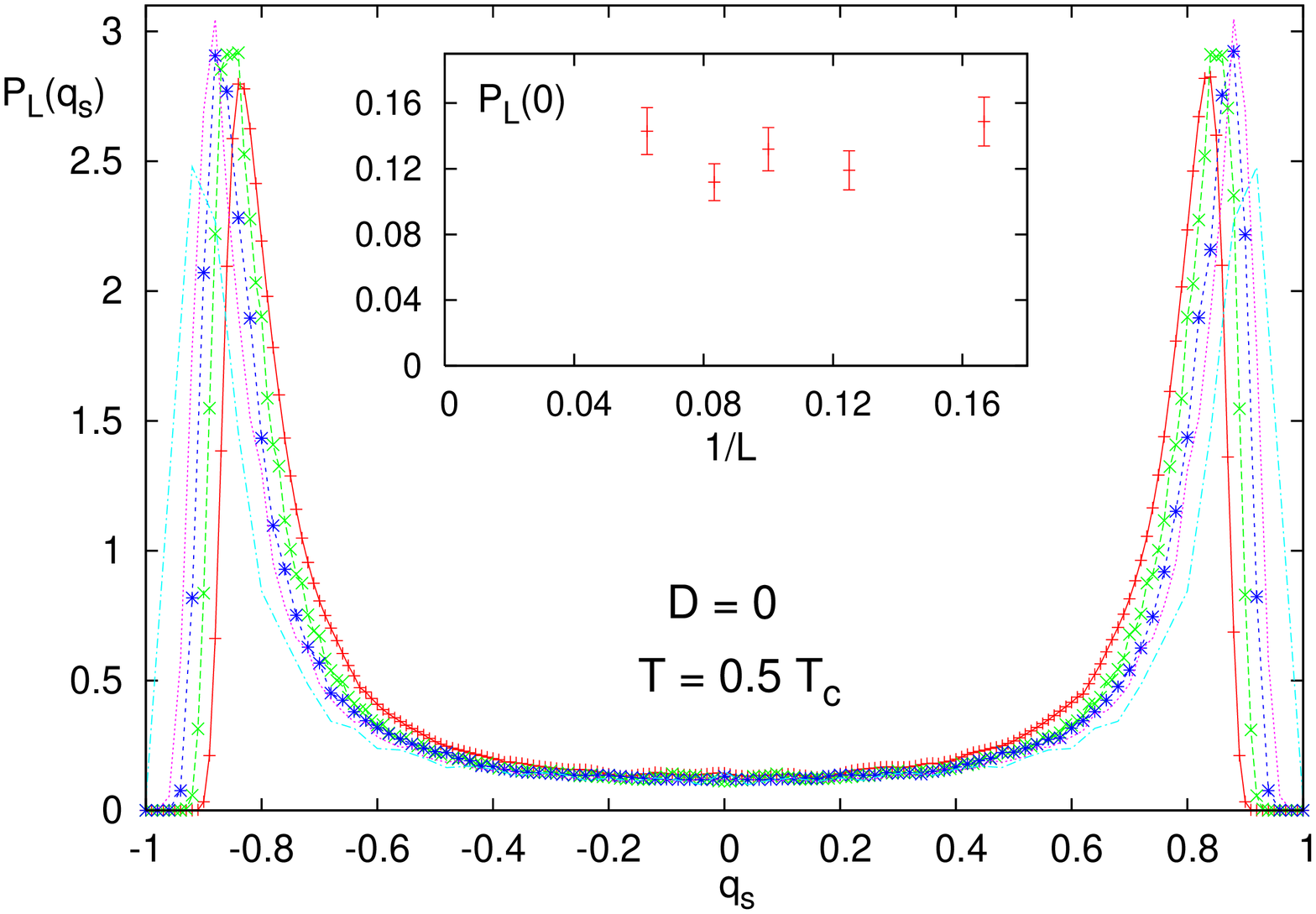}
\caption{Site overlap distribution $P_L(q_s)$ at $T=0.5$, $D=0$ for
$L=6,8,10,12,16$. Inset: $P_L(0)$ vs. $1/L$ does not tend
to zero.}
\label{f:pqs_D0_T05}
\end{figure}

\begin{figure}[!t]
\centering
\includegraphics[height=.95\columnwidth, angle=270]{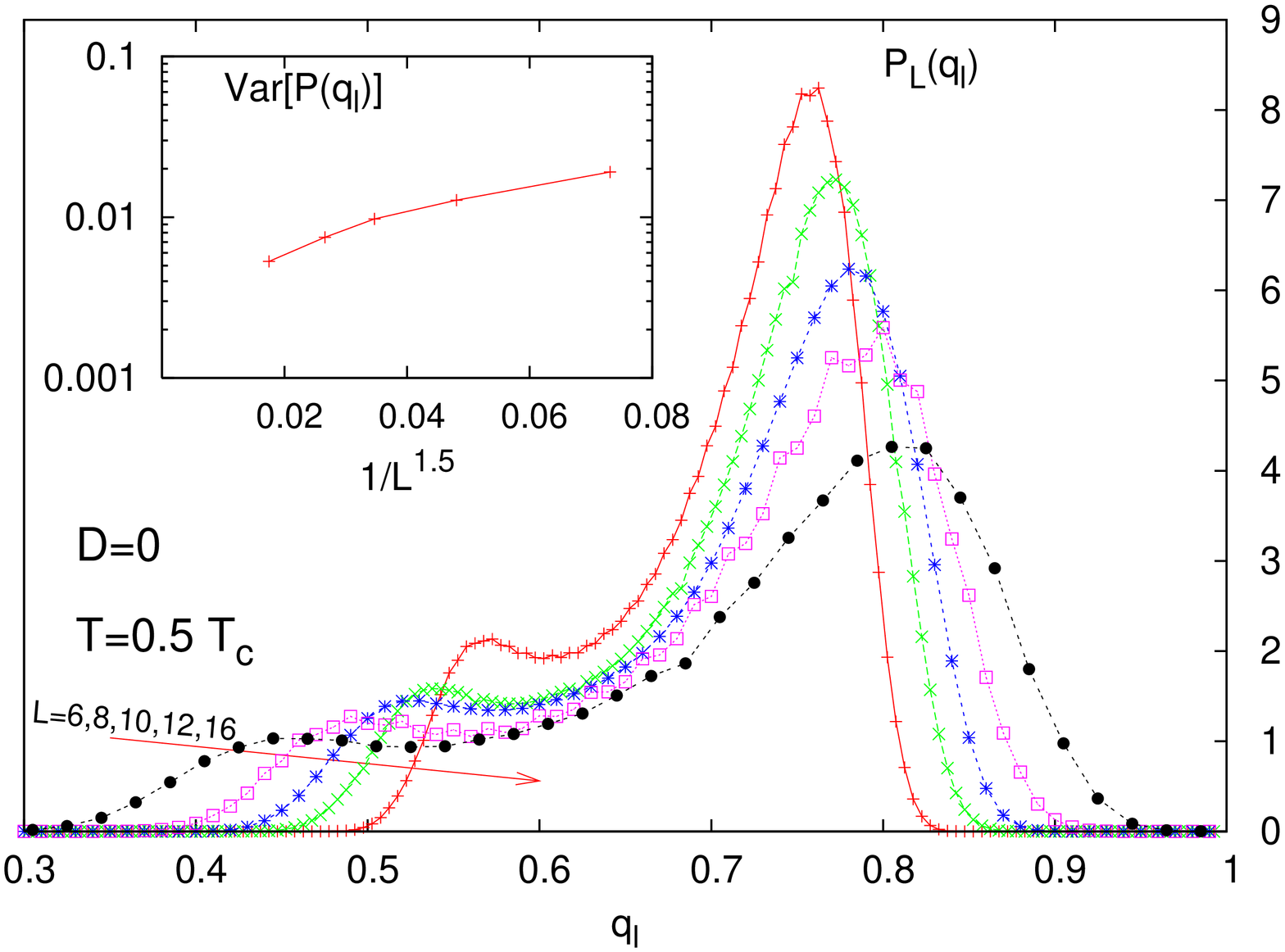}
\caption{Link overlap distribution $P_L(q_l)$ at $T=0.5$, $D=0$ for
$L=6,8,10,12,16$. Inset: Variance vs. $1/L$ tends to a very small
value $\sigma^2_{q_l}=0.0010(7)$ as $L\to\infty$ interpolating with a
power-law ($1.5(1)$).}
\label{fig:pql_D0_T05}
\end{figure}

The SG phase of the disordered BEG model, in mean-field regime, shows the same
features of the Sherrington-Kirkpatrick model:\cite{sk} in order to
obtain a stable thermodynamics the Full RSB scheme is
needed.\cite{CLPRB04,CLPRL02}  On the other hand, out of the limit of
validity of the mean-field regime, it is still unclear if the properties of SG
phase are in agreement with the RSB scenario.
The low $T,D$ phase is characterized by a pure spin-glass phase and
what this phase consists of in terms of statistical mechanic states is
the subject of the following analysis.
  Three cases are contemplated in the
literature.
\\
\indent
{\em Droplet theory:}~ it exists only one SG state (plus its symmetric
spin-reversed) and, therefore, the overlaps between states in
different replicas cannot fluctuate among different disordered samples
and the distributions are delta-shaped.\cite{droplet} The four-spins
correlation function in position space ${\bf r}=(x,y,z)$
should tend to a plateau $C_4(|r|)=q_{\rm EA}^2$, for large enough $|r|$, that
becomes longer as $T$ decreases towards $T_c$.
\\
\indent
{\em Trivial-Non-Trivial (TNT) scenario:} ~ equilibrium states are many
and non-trivially organized (i.e., $q_s$ fluctuates from sample to
sample), but the excited states are droplet-like (i.e.,
the $q_l$ overlap, sensitive to interfaces, fluctuates less and less
as the size grows). This implies that $P(q_s)$ is broad and
non-trivial, whereas $P(q_l)$ is delta-shaped.\cite{tnt} Since
excitations are trivial, the expected behavior of $C_4(x,y,z)$ is the
same as for the droplet theory.
\\
\indent
{\em Replica Symmetry Breaking (RSB) theory:} ~many states
characterize the SG phase, with space-filling excitations; both
distributions are, thus, broad, with a complex structure.
\cite{rsb,MPV86} The correlation $C_4(x,y,z)$ is expected to decay
continuously to zero (the minimum squared overlap for the present
system, in absence of an external magnetic field) at all 
$T$.\cite{Contucci07,Leuzzi08}
\\
\indent
 First we will consider the overlap distribution functions,
cf. Eqs. (\ref{f:pqs})-(\ref{f:pql}),  since, in the spin glass phase
($T<T_c$), the site and the link overlap distributions - $P(q_s)$ and $P(q_l)$ -
can be used as hallmarks to discriminate among different theories for
finite dimensional spin glasses.  In the next section we will analyze 
the four spin correlation functions. 
\\
\indent
 In order to see whether $P(q_s)$
is trivial or not we need to estimate if, for growing sizes its
support does shrink to a unique value, the Edwards-Anderson parameter $q_{\rm EA}$
or it remains finite.  In our case, in absence of an external magnetic
field, the support of a non-trivial $P(q_s)$ should range from $q_s=0$
to $q_{\rm EA}$. In Fig. \ref{f:pqs_L16_D0} we plot $P(q_s)$ at $D=0$
and size $L=16$ for all simulated temperatures: as $T$ decreases
$P(q_s)$ moves from a Gaussian to a bimodal distribution.  The
important issue is, then, whether the continuous part in between the
two peaks at low $T$ goes to zero or not as $L$ increases.  In Fig. \ref{f:pqs_D0_T05},
we plot $P(q_s)$ at the lowest thermalized temperature for
$L=6,8,10,12$ and $16$ and, in the inset, we plot the values of
$P_L(0)$ displaying no decreasing trend with increasing $L$.  The states, thus,
appear to be many and different among themselves, since they are found
with a finite probability within a non-zero continuous range of
overlap values, including $q_s=0$.
%
%\clearpage
\\
\indent
 Also $P(q_l)$ appears to develop a second peak at small $q_l$ as $L$
increases, and this signature becomes clearer and clearer at low
temperature as $L$ increases, cf. Fig. \ref{fig:pql_D0_T05}.  The
analysis of FSS of the variance of $P(q_l)$ might help to better
evaluate the breadth of the distribution in the thermodynamic
limit. Its behavior for various sizes is exemplified in the inset of
Fig. \ref{fig:pql_D0_T05}  at the lowest $T/T_c$ we
simulated for $D=0$.  The variance tends to a small finite value and we
cannot make a definitive statement about $P(q_l)$ tending towards a
delta distribution, as conjectured by the TNT scenario.  Moreover, the
study of the variance does not yield any indication about the {\em
shape} of the distribution. In particular, about the FSS behavior of
the two peaks expected in RSB theory.
\subsection{Equivalence of site and link overlap distributions}
\begin{figure}[!t]
\centering
\includegraphics[height=.95\columnwidth, angle=270]{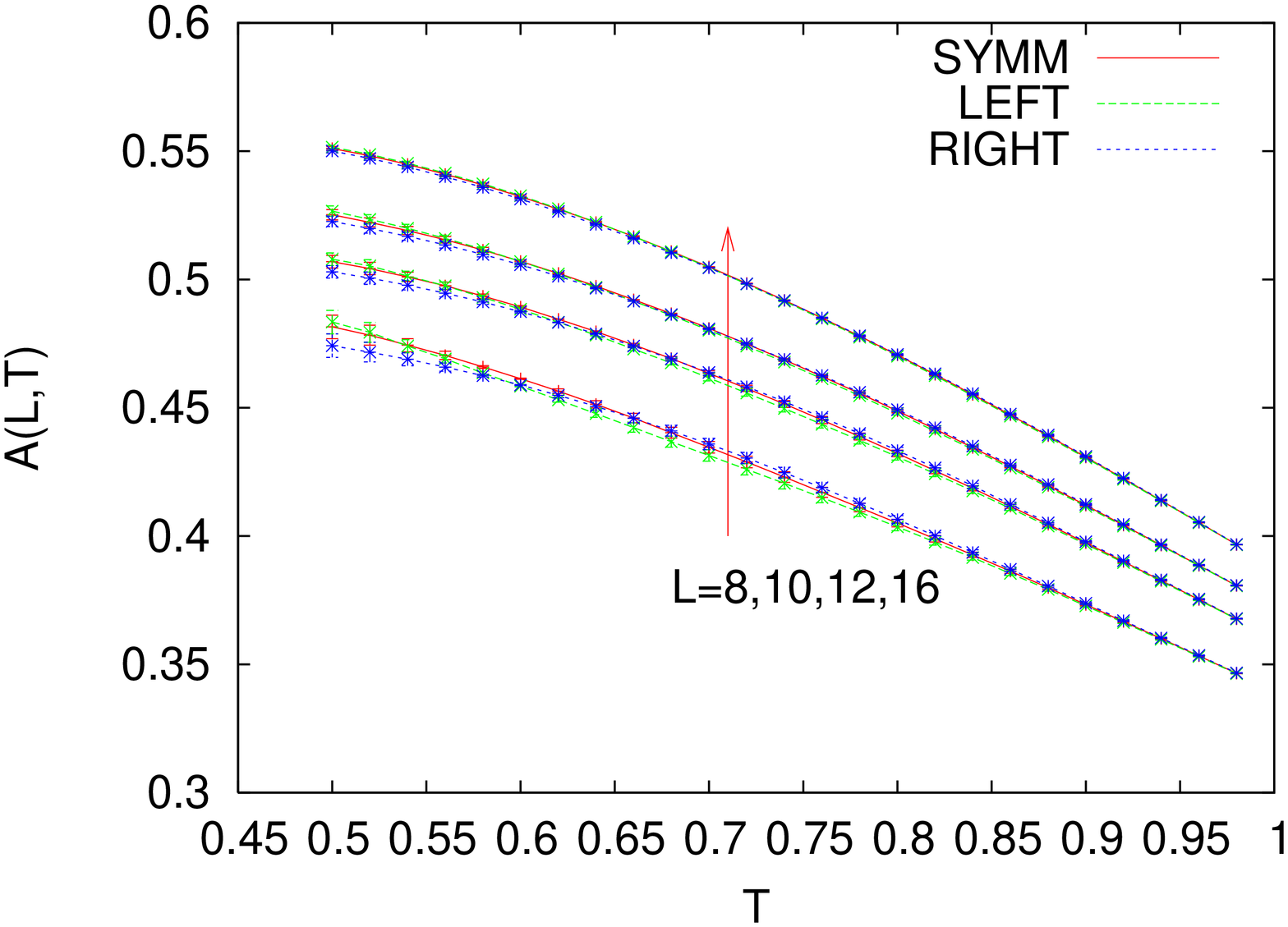}
\\
\includegraphics[height=.95\columnwidth, angle=270]{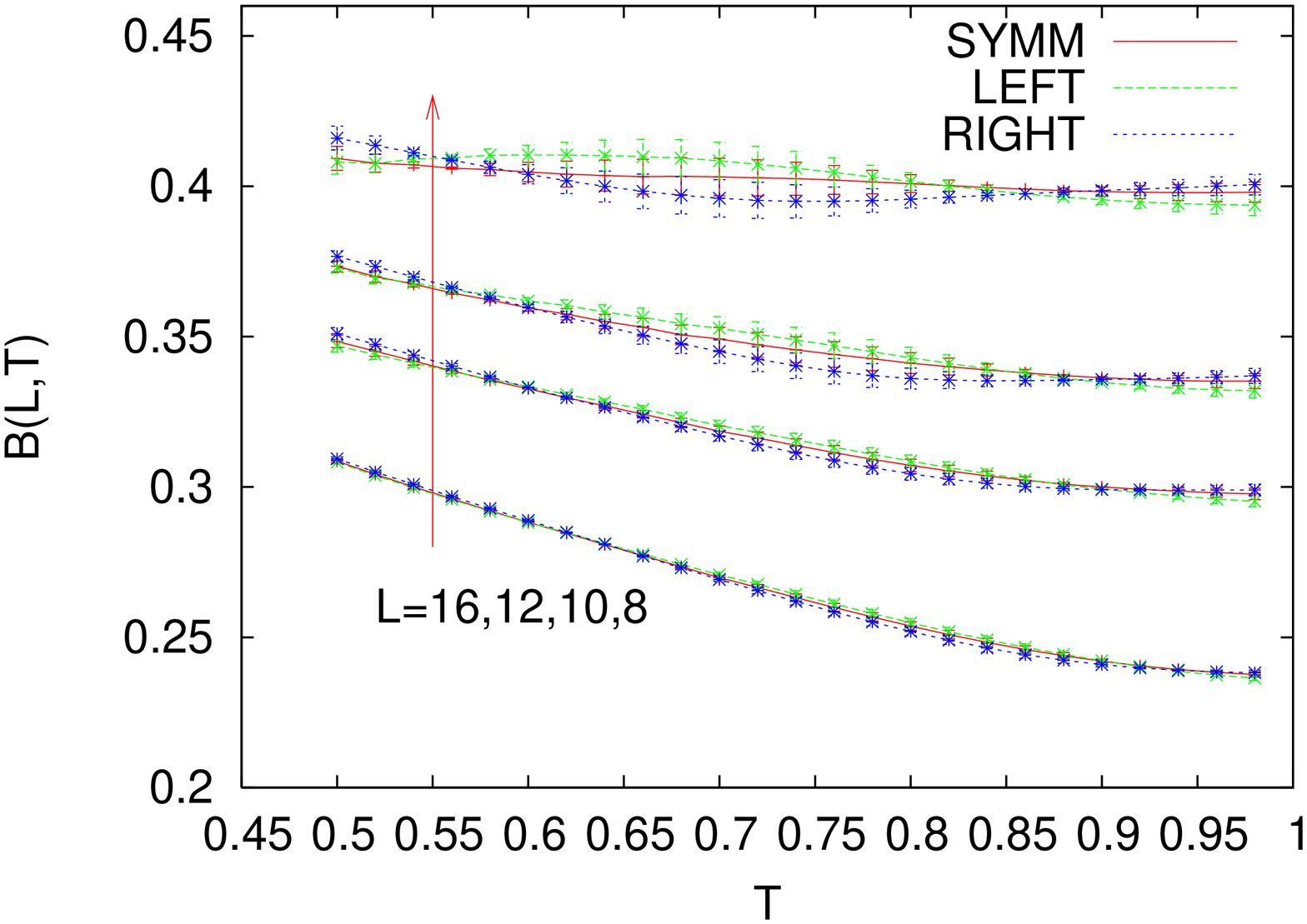}
\caption{Parameter $A$ (top) and $B$ (bottom) of $q_{\rm a}$ vs. $T$ for $L=8,10,12,16$ as
obtained minimizing the left, right and symmetric Kullback-Leibler
divergence. }
\label{fig:para_A_L}
\end{figure}

\begin{figure}[!t]
\centering
\includegraphics[width=.95\columnwidth]{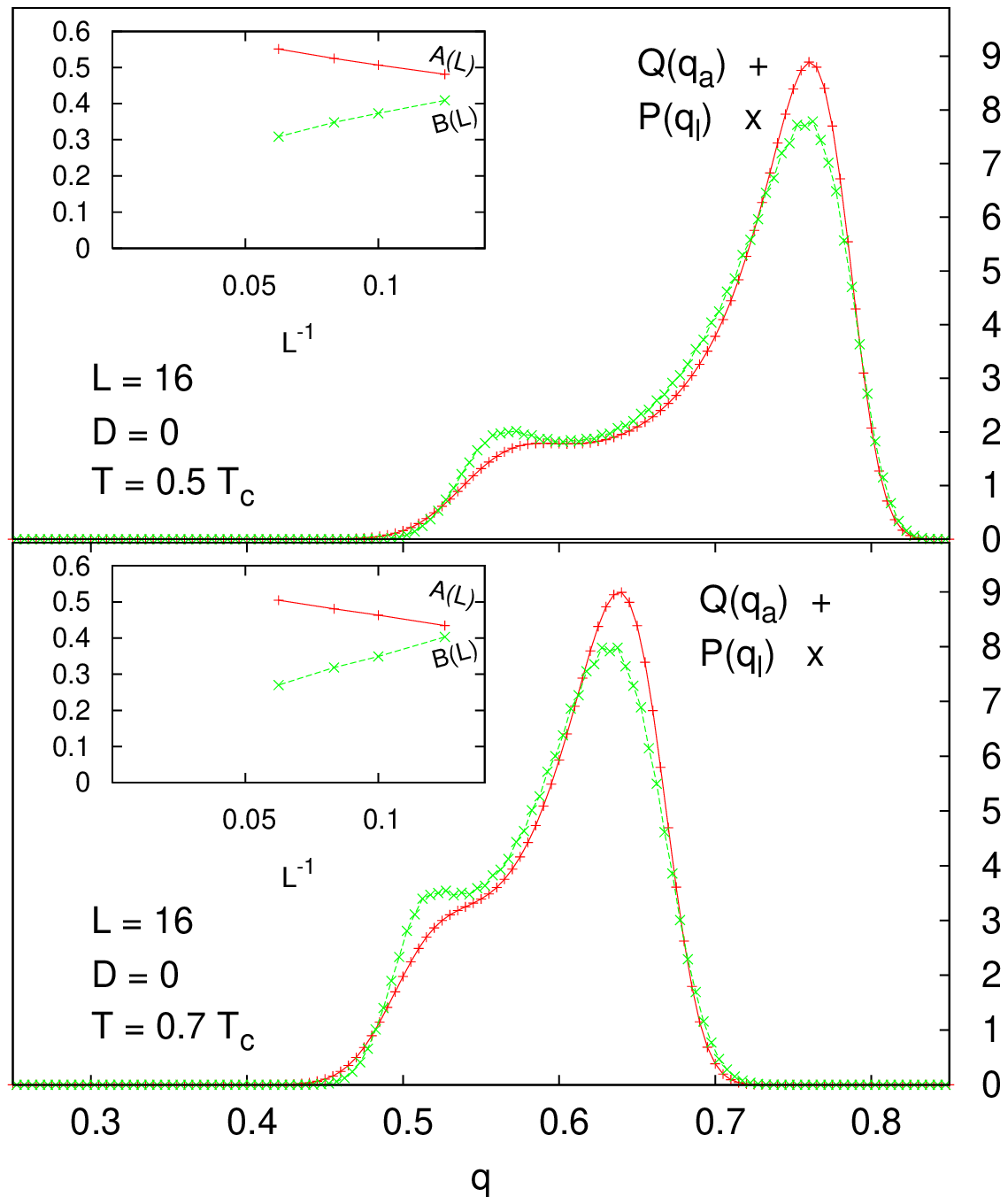}
\caption{The distributions $Q(q_a)$ and $P(q_l)$ at
$T=0.5\simeq 0.5 T_c$ (top) and $T=0.7\simeq 0.5 T_c$ (bottom), $D=0$
for an optimal choice of parameters obtained by minimizing sKLD,
cf. Eq. (\ref{f:sKLD}).  Inset: FSS behavior of the parameters $A(L)$
and $B(L)$ of the sKLD between $Q(q_a)$ and $P(q_l)$ at $D=0$ and
$T=0.5\simeq 0.5 T_c$. Sizes are for $L=6,8,10,12,16$.  }
\label{fig:PqQq_L16_T05}
\end{figure}

\begin{figure}[!t]
\centering
\includegraphics[height=.95\columnwidth, angle=270]{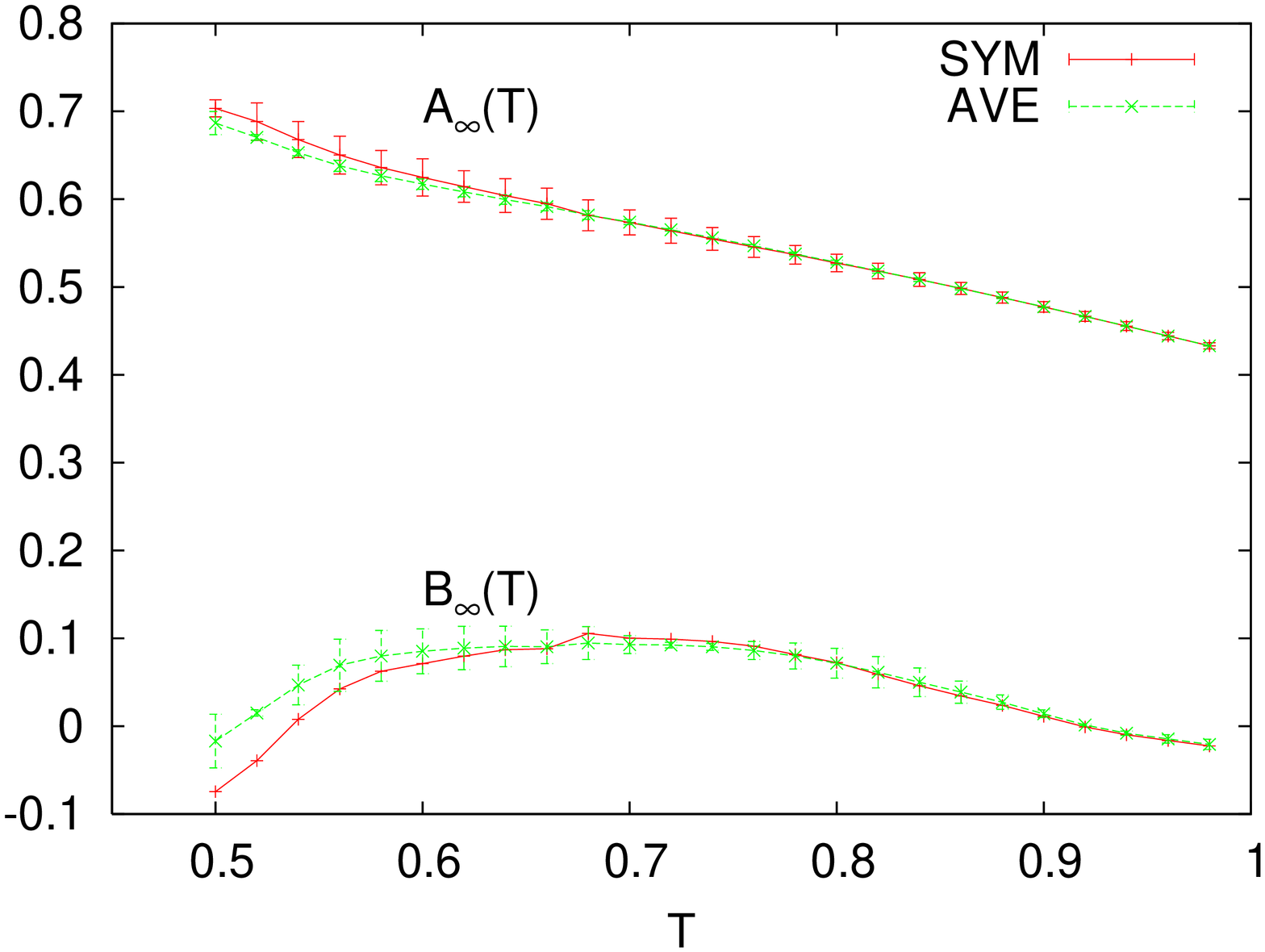}
\caption{FSS limit $L\to \infty$ of $A$ and $B$ parameters vs. $T$.}
\label{fig:para_A_inf}
\end{figure}

We can, then, implement a more refined analysis of the pdf data
and check whether $P(q_s)$ and $P(q_l)$ are actually
equivalent and, thus, if the non-triviality of the former implies the
non-triviality of the latter.  This can be realized by recalling that
in the SK model $q_l=q_s^2$ and by comparing $P(q_l)$ to the
distribution $Q(q_a)$ of an auxiliary variable
\BEQ
q_{a}\equiv A+B q_{s}^2 + z\sqrt{1-q_{s}^2}
\EEQ
 with $z$ a Gaussian random variable of variance $\sigma_z$ and zero
mean, that mimics the presence of fluctuations due to the finite size
of the considered systems.
\\ \indent At a given point of the phase diagram $D,T$ and for a given
size $L$, the parameters $A(L)$, $B(L)$ and $\sigma_z(L)$ can be
obtained by minimizing the 
 Kullback-Leibler divergence\cite{KL} (KLD) between $P(q_{l})$ and $Q(q_{a})$:
\BEQ
D_{\rm KL}[P,Q]= \sum_{i=1}^{N_{\rm bin}}
P(q_i)\log\frac{P(q_i)}{Q(q_i)}
\EEQ
We will refer to this one as the ``left'' KLD. The ``right'' KLD is the same formula exchanging
$P$ and $Q$, where the  symmetrized divergence
(sKLD) between $P(q_{l})$ and $Q(q_{a})$ is defined as: \cite{Leuzzi08}
\BEQ
D_{\rm KL}[P,Q]= \frac{1}{2}\sum_{i=1}^{N_{\rm bin}}\left[
P(q_i)\log\frac{P(q_i)}{Q(q_i)}+Q(q_i)\log\frac{Q(q_i)}{P(q_i)}
\right]
\label{f:sKLD}
\EEQ
In Fig.  \ref{fig:para_A_L} we plot, the finite size values of the
 parameters $A$ and $B$. Besides the values of the parameters
 minimizing the symmetrized KLD, Eq. (\ref{f:sKLD}) we also plot the
 values of $A$ and $B$ minimizing the left and the right unsymmetrized
 KLD's. We observe that, as $L$ increases the spread between different
 estimates tends to vanish. The infinite size limit of $\sigma_z$ is
 always compatible with zero, signaling that FS effects actually tend to vanish as $L$ increases,  though with
 large statistical errors at low temperature, implying that smaller sizes might hinder a correct FSS. 
\\ \indent As instances we plot the matching of the two distributions
$Q(q_{a})$ and $P(q_l)$ in Figs.  \ref{fig:PqQq_L16_T05} at
$T=0.5\simeq 0.5 T_c$ and $T=0.7\simeq 0.7 T_c$ at size $L=16$ and
$D=0$.
 In the insets we plot the size behavior of $A$ and $B$ from the sKLD
for the two specific cases.  In the first case, performing a power-law
FSS scaling to $L\to \infty$ we obtain that $B$ interpolates a
negative value!  In the second case the $L\to\infty$ limit yields a
positive value.  This observation is contrasting from the behavior,
cf. bottom panel of Fig. \ref{fig:para_A_L}, of $B(T)$ growing with
decreasing $T$ at all fixed sizes. Quite evidently, the low $L$ strong
fluctuations strongly bias the interpolation at small $T$. To show it
in a clearer way, in Fig. \ref{fig:para_A_inf} we plot the asymptotic
values of both $A$ and $B$ for all simulated temperatures both from the sKLD and as the average of
the extrapolation of the values minimizing the right and left
unsymmetrized KLD's.  With $A_\infty(T)$ the two estimates appear to
be consistent at all temperature and reproduce the qualitative
behavior detected in all finite $L$ cases, compare with
Fig. \ref{fig:para_A_L}. For $B_\infty(T)$, at low $T$ the two
estimates are not consistent anymore.  Moreover, $B_\infty(T)$
decreases with $T$ below a certain $T\simeq 0.7$, unlike its finite
$L$ counterparts (at least as $L\geq 10$),
cf. Fig. \ref{fig:para_A_L}.
\\ \indent We face strong finite size effects and a crossover between
small and large sizes is taking place. However, due to the fact that
we cannot easily thermalize larger systems at low temperature, we
cannot make any definite statement on the behavior of $B_\infty(T)$
for very low $T$.  We simply do not have enough reliable points in $L$
at our disposal.
The finite size behaviors, though, strongly suggest that $Q(q_a)$ and
$P(q_l)$ are, indeed, equivalent even below $T=0.7$. In any case, the
equivalence is proven for $T\geq 0.7$ implying that not only the
equilibrium states have a non-trivial distribution but also their
excitations, yielding evidence in favor of the third scenario
considered, the RSB theory, rigorously valid in mean-field systems.

\subsection{Position Space Four Spins Correlations}
\begin{figure}[t!]
\centering
\includegraphics[width=.95\columnwidth]{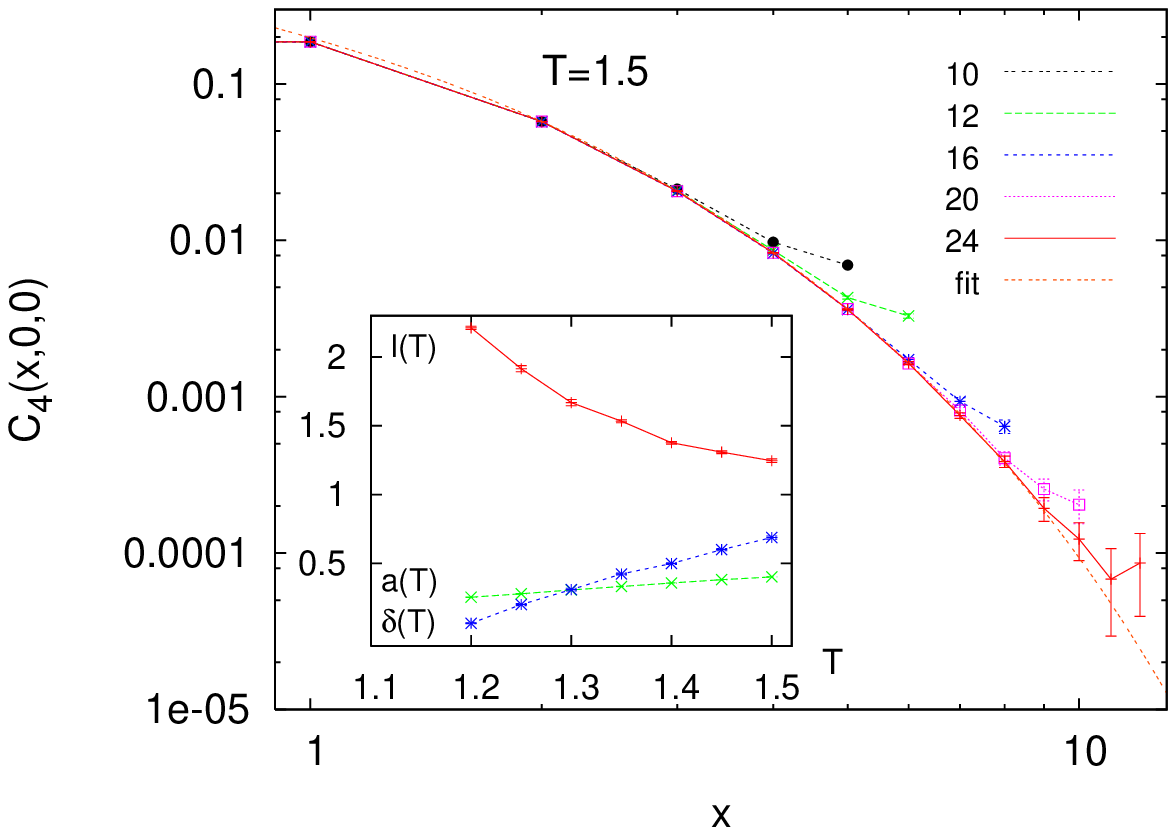}
\caption{Correlation between local overlap for $D=0$ and sizes
$L=10,12,16,20,24$, at the largest simulated temperature $T=1.5$.  The
 fit with Eq. (\ref{f:C4fit_hT}) is also plotted. }
\label{fig:C4hT}
\end{figure}

\begin{figure}[t!]
\centering
\includegraphics[width=.95\columnwidth]{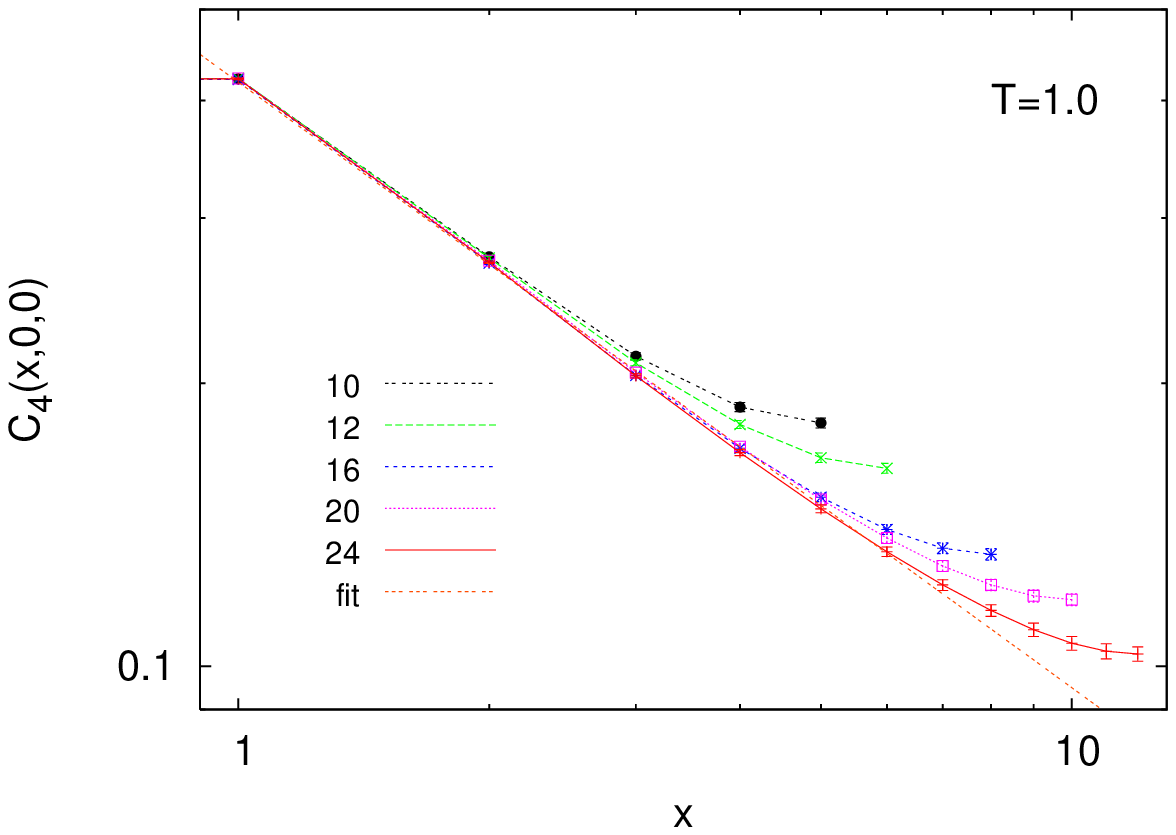}
\caption{Behavior of $C_4(x)$ at $D=0$, for $L=10,12,16,20,24$ and
$T=1\simeq T_c$. The interpolation with a simple power-law,
$\alpha=0.64(1)$, is shown for $L=24$. On shorter systems:
$\alpha=0.64(1), L=20$ and $\alpha=0.65(2), L=16$.}
\label{fig:C4Tc}
\end{figure}

\begin{figure}[t!]
\centering
\includegraphics[width=.95\columnwidth]{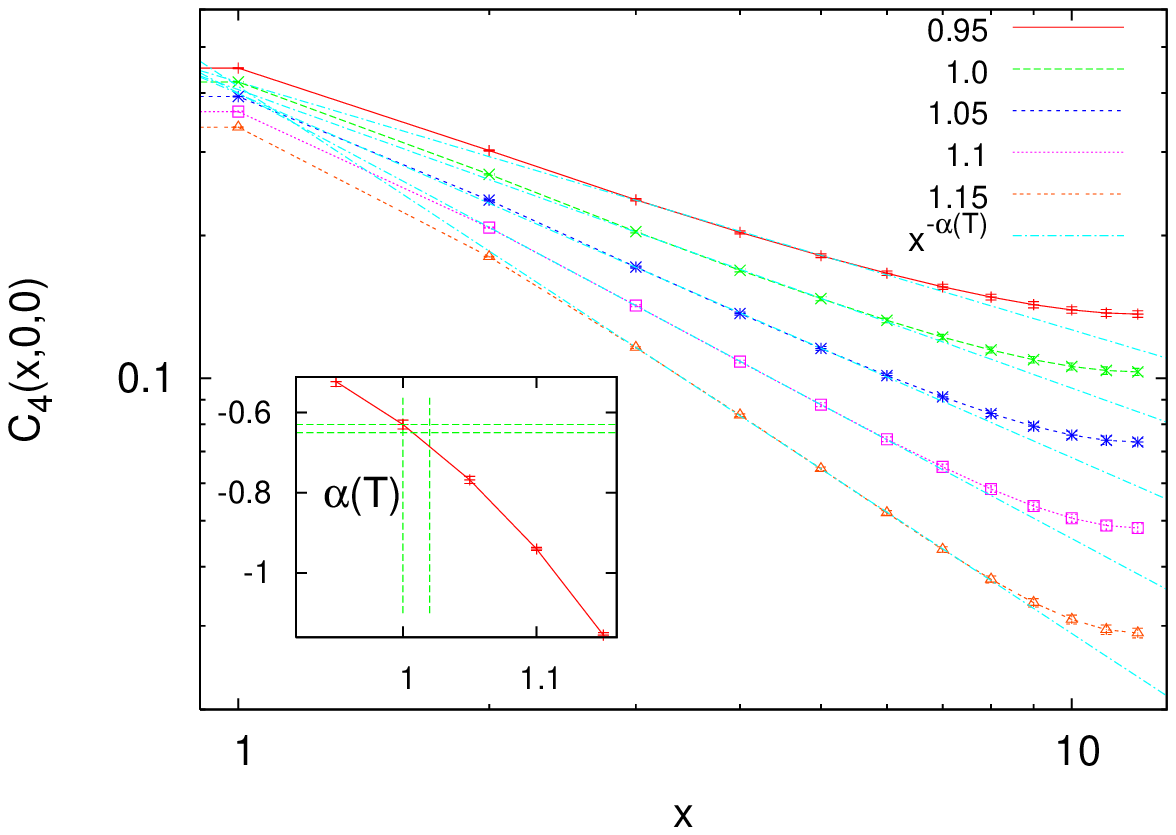}
\caption{Behavior of $C_4(x)$ for $L=24$ and $T=0.95,1,1.05,1,1.15$.
The interpolation with a simple power-law is also shown for $L=24$. Inset:
behavior of the power $\alpha$ vs. T. The dashed vertical and
horizontal lines denote, respectively the estimates of $-d+2-\eta$ and
$T_c$ (with errors, cf. Tab. \ref{tab:crit}: $T_c=1.01(1)$, $\eta=-0.35(1)$). }
\label{fig:C4lT}
\end{figure}

We now investigate the behavior of the four spins correlation
function, defined in Eq. (\ref{f:C4def}), in position space.  We
recall that the droplet and TNT theories predict that $C_4(x)$ tends
to a plateau of height $q_{\rm EA}^2$ (cf. Sec. \ref{sec:theories})
whereas RSB theory predicts for $C_4(x)$ at $T<T_c$ a power-law decay
$\sim x^{-\alpha}$.  We, thus, have to compare our data with the
prediction of one of these hypotheses.
\\ \indent Since we are dealing with small systems, we must first
consider possible FS effects.  Indeed, because of the periodic
boundary conditions imposed on the simulated system, the correlation
function that we actually measure at a distance $x$ also contains the
contribution of correlations at distance $x+kL$, with
$k=1,\ldots,\infty$ and the true (yet unknown) correlation function
${\cal C}_4(x,y,z)$ is related to the measured one - $C_4(x,y,z)$ - by
the relationship:
 \BEQ C_4(x,y,z)= \sum_{k_x,k_y,k_z}^{0,\infty} {\cal
 C}_4(x+k_xL,y+k_yL,z+k_zL) \EEQ
For large distances, when $C_4$ is smaller, these extra contribution
will strongly bias the estimate of the true ${\cal C}_4$ behavior in
space. In particular, correlations at larger distances, of order $L/2$, 
 will experience relatively 
stronger systematic errors than $C_4(|r|\ll L)$.
\\ \indent We will now present our results for the case $D=0$.  For
temperatures down to the critical region we simulated lattices with
sides of length up to $L=24$.  The largest thermalized size for $T$
down to $0.5 T_c$ is, instead, $L=16$.  In Fig.  \ref{fig:C4hT} we
plot the $x$ behavior at $T=1.5$ in a log-log plot for the sizes
$10,12,16,20,24$. One can observe that FS effects are limited to the
last point at $L/2$. The rest of the curves completely superimpose.
\\
\indent
At high temperature, correlations are expected to decay exponentially
at large enough distances. As temperature is lowered towards
criticality the $C_4(x)$ should become power-law eventually decaying
as $x^{-d+2-\eta}$ at $T=T_c$.
We, then, interpolate the four-spins correlation function along the
$x$-axis at criticality with the function:
\BEQ C_4^{\rm fit}(x)= a
x^{-\alpha}\left[1+\left(\frac{x}{\ell}\right)^{-\delta\alpha}e^{\delta
x/\ell}\right]^{-1/\delta}
\label{f:C4fit_hT}\EEQ
and equivalently for $y$ and $z$, due to the anisotropy of the system
in absence of an external field.  This is a function containing a
crossover between a short distance power-law decay, $x^{-\alpha}$, and
an exponential decay, with characteristic 'correlation' length $\ell$.
In Fig.  \ref{fig:C4hT} the function interpolating the $L=24$
$C_4(x,0,0)$ is plotted with $a=0.402(9)$,
$\delta=0.69(1)$,$\ell=1.25(1)$ with $\chi^2=0.088$.  As the
temperature decreases the correlation length increases until it
becomes too long to be observed in the analyzed systems.  In the inset
of Fig.  \ref{fig:C4hT} we plot the $T$ behavior of $\ell$, $\alpha$
and $\delta$ until the fit becomes inconsistent $T\simeq 1.15$.
\\ \indent In Fig. \ref{fig:C4Tc} we plot the $C_4$ curves at $T\simeq
T_c$ for sizes $L=10,12,16,20,24$, as well as the interpolation of the
latter with $Ax^{-\alpha}$ (the correlation length is too long to
detect the exponential contribution in Eq. (\ref{f:C4fit_hT})). The
exponent equals the power at criticality $\alpha=d - 2+\eta =0.64(1)$
(at crystal field $D=0$ it was $\eta=0.36(1)$,
cf. Tab. \ref{tab:qm}). 
\noindent At $T=1$ the interpolated value of $\alpha$
for the $L=24$ $C_4(x)$ curve is $\alpha = 0.64(1)$, $\alpha=0.65(2)$ for
$L=16$ and  $\alpha=0.64(1)$ for $ L=20$.  FS effects appear to be stronger now
w.r.t. Fig. \ref{fig:C4hT} and evident also for $x<L/2$ (only points
for $x\leq L/4$ actually stay on the $x^{-\alpha}$ curve).
\\
\indent
Approaching $T_c$, as $T<1.2$, cf. Fig. \ref{fig:C4lT}, it is
not possible to detect a crossover between power-law and exponential
decay and the simple power-law decay is tested. In the inset the power
behavior in $T$ is shown and compared with the power at criticality,
$\alpha=0.65(1)$. 
\begin{figure}[!t]
\centering
\includegraphics[width=.95\columnwidth]{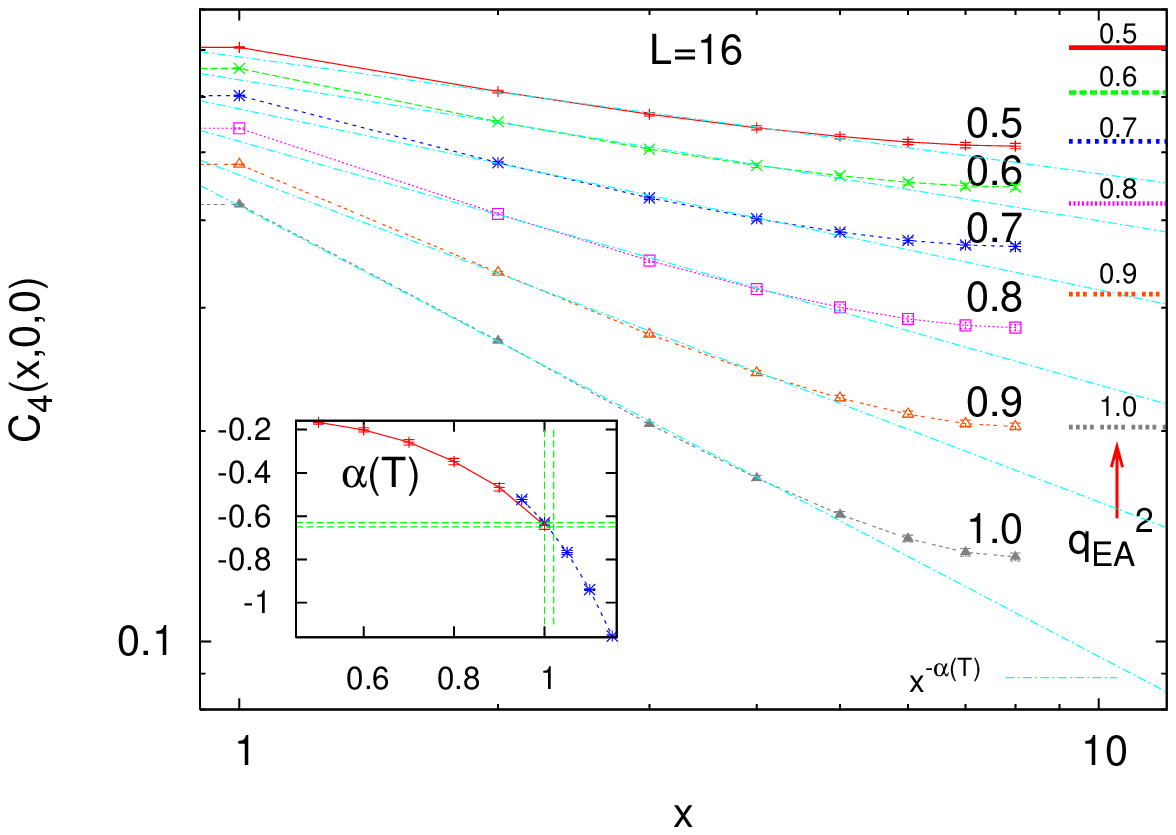}
\caption{Behavior of $C_4(x)$ for $L=16$ and
$T=0.5,0.6,0.7,0.8,0.9.1.0$.  The interpolation with a simple
power-law is also shown. On the right hand side the values of $q_{\rm
EA}^2(T)$ are displayed. Inset: behavior of the power $\alpha$ vs. $T$
(full line) compared to the $\alpha(T)$ behavior around criticality
for $L=24$ (dotted line).  }
\label{fig:C4lT16}
\end{figure}
\\
\indent
Decreasing further the temperature we show in Fig. \ref{fig:C4lT16}
that the behavior is power-law until $x\sim L/4$ is reached. At that
point the curves bend upwards as it did at criticality and even at
high temperature, cf. Fig. \ref{fig:C4hT}.  This bending is, however,
an artifact due to the contributions induced by the periodic boundary
conditions.  In Fig.  \ref{fig:C4lT16}, on the right hand side, we
show the values of $q_{\rm EA}^2$ at the same temperatures of the
plotted $C_4(x)$. At all temperatures the $C_4(x)$ soon decays below
the corresponding value of $q_{\rm EA}^2$.  For the sizes simulated
our data are, thus, not consistent with the observation of a plateau
at $q_{\rm EA}^2$ as predicted by the droplet and TNT theories.

\section{Conclusions}

In the present work we have performed Parallel Tempering Monte Carlo
numerical simulations in the temperature/crystal field plane of the random 3D
Blume-Capel model on a cubic lattice. This is a spin-1 spin glass,
whose constituent features try to capture at least one supposed
mechanism underlying inverse transitions: the raise of inactive
components at low $T$.

In particular, we have analyzed the second order phase transition
carrying out the computation of the critical temperatures and indeces
by means of parallel tempering simulations in temperature at different
values of the chemical potential $D$. In this analysis we have
carefully checked FS effects, identified eventual crossovers from
small to large size scaling and neglected data for correspondingly too
small sizes.  We verified that for different values of $D$ the system
is always in the same universality class (as far as a continuous
transition occurs) looking, e.g., at different universal scaling
functions of $\xi_c/L$, such as the Binder parameter $g$, or the
quotients of $\chi_{\rm SG}$, $\xi_c$ and $g$ between systems at $L$ and $2L$.
The outcome is that at all $D<D_{3c}$ the second order transition
belongs to the same universality class of the 3D Edwards-Anderson
model for spin-glasses. 
\\
\indent
We, then, estimated the position of a tricritical point, $D\sim 2.1$,
 $T\sim 0.5$, beyond which the transition is first order with jump in
 density and in overlap parameters. This transition is first order in
 the thermodynamic sense, i.e., latent heat is exchanged and, even
 though the system is disordered, it is not related to the random
 first order transition taking place in structural glasses.
 \cite{LeuzziNieuw07}  We employed and compared four different
 methods to infer the critical line from FS data. 
 This  observation confirms the claim
of Fernandez {\em et al.} \cite{Fernandez08} about the existence of
such transitions in quenched disordered short-range finite-dimensional
systems.  In the present model the first order transition can be seen by
means of standard parallel tempering algorithm in the canonical ensemble,
simply tuning an external pressure-like parameter. 
\\ \indent The first order transition line has the property of
displaying inverse freezing, as can be observed from the phase
diagram, cf. Figs. \ref{fig:phdi}, \ref{fig:phdi_det}: the low
temperature phase is paramagnetic and the system 'freezes' into a
spin-glass phase as $T$ is {\em increased}. This is at difference with
the thermodynamic behavior of the original, ordered, BC model
(mean-field or finite dimensional). \cite{CapelPhys66,Saul74} In
presence of quenched disorder, a low temperature paramagnetic phase
exists that can acquire a very low density and this is the source of the
entropy decrease with respect to the high temperature paramagnetic
phase.
\\
\indent
Both the inverse freezing transition and its first order nature
 were not observed in the same model
on a hierarchical lattice. \cite{Ozcelik08}
\\
\indent
Eventually we present our analysis of the overlap distribution
functions and the four-spins correlation functions at criticality and
in the glassy phase, at $D=0$ for $T$ down to $0.5 T_c$.  From the
behavior of site overlap distribution at zero overlap, $P_L(q_s=0)$,
and from the variance of the link-overlap distribution $P_L(q_l)$ we
get evidence in favor of a complex organization of states in the SG
phase, displaying features typical of the Replica Symmetry Breaking
theory holding for mean-field systems ($d\geq 6$). We cross-checked
this observation comparing, with the Kullback-Leibler divergence, the
link-overlap distribution with the distribution of a function of the
squared site overlap, $q_a\sim A + B q_s^2$. We carefully analyzed the
finite size effects at low temperature finding that for $T<0.7$ small
size fluctuations strongly bias our estimates, yielding {\em negative}
$B$ coefficients of the $q_s^2$ term, {\em decreasing} with
temperature, unlike any finite size $B(T)$ behavior. In order to have
a self-consistent estimate we would need to thermalize at $T\geq
0.7T_c$ systems of size sensitively larger than $L=16$.
\\
\indent
Looking at the position dependence of the four-spins correlation 
functions we are able to detect, for $T\geq 1.2T_c$, a crossover
between a short-distance power-law decay and a long-distance
exponential decay and we can identify a length-like parameter $\ell$
playing the role of the correlation distance, growing as $T$
decreases.  As the critical temperature is approached and $\ell$
becomes similar to the maximum feasible distance in the simulated
system ($\sim L/2$), $C_4(x)$ can be interpolated with a simple
power-law. We checked that for sizes $L=16, 20$ and $24$ the exponent
of $C_4(x)$ at $T_c$ is equal to $d-2+\eta$, where $\eta=-0.36(1)$ is the value
obtained from the analysis of the critical properties performed with
the quotient method.  We also probed the power-law behavior for temperatures
down to $0.5 T_c$ at distances far away from border, where finite size correction
are too strong. Indeed, periodic boundary conditions systematically increase
correlations, above all where they are small (or vanishing), i.e., at large distance. We
compare the low temperature behavior with the prediction of TNT and
droplet theories that $C_4(x)$ should tend to a plateau $C_4\sim
q_{\rm EA}^2$ for large $x$. Even though we are not able to reach
``large $x$'', we show that $C_4(x) < q_{\rm EA}^2$ already at small
distance.

\acknowledgments
We thank Nihat Berker, Helmut Katzgraber and Federico Ricci-Tersenghi 
for interesting discussions and exchanges.


\begin{thebibliography}{99}
\bibitem{Tammann} G. Tammann, ``Kristallisieren und Schmelzen'',
Metzger und Wittig, Leipzig (1903).

\bibitem{Wilks87}J. Wilks, D.S. Betts, {\em
An Introduction to Liquid Helium}, Oxford University Press (USA, 1987).

\bibitem{RHKMM99} S. Rastogi, G.W.H. H{\"o}hne and A. Keller,
Macromolecules {\bf 32}, 8897 (1999).

\bibitem{Greer00}
 A.L. Greer, Nature {\bf 404},
134 (2000).

\bibitem{vRRMM04} N.J.L. van Ruth and S. Rastogi, Macromolecules {\bf 37},
8191 (2004).

\bibitem{Plazanet04}
M. Plazanet {\em et al.}
J. Chem. Phys. {\bf 121}, 5031 (2004).

\bibitem{Tombari05} E. Tombari {\em et al.}, J. Chem. Phys. {\bf 123}, 051104 (2005).

\bibitem{Plazanet06}
M. Plazanet {\em et al.}
%, M. Dean, M. Merlini, A. Huller,
%H. Emerich, C. Meneghini, M. R. Johnson, and
%H. P. Trommsdorff.
J. Chem. Phys. {\bf 125}, 154504 (2006).

 \bibitem{Plazanet06b}
 M. Plazanet {\em et al.}, Chem. Phys.  {\bf 331}, 35 (2006).

\bibitem{Angelini07}
R. Angelini and G. Ruocco, Phil. Mag.  {\bf 87}, 553 (2007).

\bibitem{Ferrari07} C. Ferrari {\em et al.}, J. Chem. Phys.
{\bf 126}, 124506 (2007).

\bibitem{Angelini08}
R. Angelini, G. Salvi  and G. Ruocco, Phil. Mag.  {\bf 88}, 4109 (2008).

\bibitem{Angelini08b}
R. Angelini, G. Ruocco, S. De Panfilis, Phys. Rev. E {\bf 78}, 020502 (2008).

\bibitem{Plazanet09}
 M. Plazanet, M.R. Johnson and H.P. Trommsdorff, Phys. Rev. E {\bf 79},
 053501 (2009).

 \bibitem{Angelini09}
R. Angelini, G. Ruocco and S. De Panfilis, Phys. Rev. E {\bf 79}, 053502 (2009).

\bibitem{CACPS97} C. Chevillard and M.A.V.  Axelos,
Colloid. Polym. Sci. {\bf 275}, 537 (1997).

\bibitem{Hirrien98}
M. Hirrien {\em et al.},
Polymer {\bf 39}, 6251 (1998).

\bibitem{Haque93}
A. Haque and E.R. Morris,  Carb. Pol. {\bf 22}, 161 (1993).

\bibitem{Avraham01}    N. Avraham {\em et al.},  Nature  {\bf 411} 451 (2001).

\bibitem{Greiner02} M. Greiner {\em et al.},
% O. Mandel, T. Esslinger, T.W. Hänsch,
%and I. Bloch,
Nature {\bf 415} 39 (2002).

\bibitem{Donnio07} B. Donnio {\em et al.},
% P. García-Vázquez, J.-L. Gallani,
%D. Guillon, and E. Terazzi,
Adv. Mater. {\bf 19}, 3534 (2007).

\bibitem{Donnio10} B. Donnio {\em et al.},
Soft Matter {\bf 6},  965 (2010).

\bibitem{Zaccarelli02}
E. Zaccarelli {\em et al.}, Phys. Rev. E {\bf 63},
031501 (2002).

\bibitem{Zaccarelli04} E. Zaccarelli {\em et al.}
 Phys. Rev. E {\bf 66}, 04102 (2004).

\bibitem{Cladis75}
P.E. Cladis, Phys. Rev. Lett. \textbf{35}, 48 (1975);

\bibitem{Cladis77}
P.E. Cladis, Phys. Rev. Lett.
\textbf{39}, 720
(1977);

\bibitem{Ozbek02}
H. {\"O}zbek {\em et al.}, Ph. Trans. {\bf 75}, 301 (2002);

\bibitem{Portmann03}
O. Portmann, A. Vaterlaus, and D. Pescia,  Nature {\bf 422}, 701 (2003).

\bibitem{Srivastava07} A. Srivastava, D. Sa and S. Singh, Eur. Phys. J. E {\bf 22}, 111
(2007);

 \bibitem{Jaffar97}
 B. M. Jaffar
Ali and A. Kumar, J. Chem. Phys. {\bf 107}, 8020 (1997).

\bibitem{Bagchi06}
 D. Bagchi,
A.Kumar and R. Menon, J. Chem. Phys. {\bf 125}, 034511 (2006).

\bibitem{Verbeek78}
H. Verbeek, G.J. Nieuwenhuys, H. Stocker, and
J.A. Mydosh, Phys. Rev. Lett.  {\bf 40}, 586 (1978).

\bibitem{Yeshurun80}
Y. Yeshurun,
M.B. Salamon, K.V. Rao, and H.S. Chen, Phys. Rev. Lett.  {\bf 45},
1366 (1980).

\bibitem{SDTJPC01} F.H. Stillinger, P.G. Debenedetti and T.M. Truskett,
J. Phys. Chem. B {\bf 105}, 11809 (2001).

\bibitem{SDBBPC03}
F.H Stillinger and P.G. Debenedetti, Biophys. Chem.
{\bf 105}, 211 (2003).

\bibitem{Feeney}
M.R. Feeney., P.G. Debenedetti, and F.H. Stillinger,
 J. Chem. Phys. {\bf 119} 4582 (2003).

 \bibitem{SSPRL04}
  N. Schupper and N.M. Shnerb, Phys. Rev. Lett.  {\bf 93} (2004)
  037202.

\bibitem{SS05} N. Schupper and N.M. Shnerb,  Phys. Rev. E, 72: 046107, 2005.

\bibitem{Prestipino}
    S. Prestipino, Phys. Rev. E {\bf 75}, 011107 (2007).

\bibitem{CLPRL05}  A. Crisanti and L. Leuzzi, Phys. Rev. Lett. {\bf 95},
 08720170 (2005).

\bibitem{Sellitto06}M. Sellitto,  Phys. Rev. B {\bf 73} 180202 (2006).

\bibitem{SKPRL05} M. Sellitto and J. Kurchan, Phys. Rev. Lett. {\bf 95},
236001 (2005).

\bibitem{APPRL06} A. Allahverdyan and Petrosyan, Phys. Rev. Lett. {\bf 96},
065701 (2006).

\bibitem{CapelPhys66}H. W. Capel, Physica {\bf{32}}(1966) 966;
M. Blume, Phys. Rev. {\bf{141}} (1966) 517.

\bibitem{GSJPC77} S. K. Ghatak, D. Sherrington, J. Phys. C: Solid State Phys.
{\bf{10}}, 3149 (1977).

\bibitem{Dumas} J. Dumas {\em et al.},
% C. Schhlenker, J. L. Tholence and
%R. Tournier
Phys. Rev. B {\bf 20}, 3913 (1979).
%``AIP Conf. Magnetism and Magnetic Materials,
%Philadelphia''.

\bibitem{CLPRL02} A. Crisanti and  L. Leuzzi,
Phys. Rev. Lett. {\bf 89} (2002) 237204.

\bibitem{CLPRB04} A. Crisanti and L. Leuzzi, Phys. Rev. B {\bf 70}
(2004) 014409.

\bibitem{LPM} L. Leuzzi, \emph{Phil. Mag.} \textbf{87}, 543-551
(2006).

\bibitem{BEGPRA71} M. Blume, V.J. Emery and R.B. Griffiths,
 Phys. Rev. A {\bf{4}} (1971) 1071.

\bibitem{Saul74} D.M. Saul, M. Wortis and D. Stauffer, Phys. Rev. B {\bf  9}, 4964 (1974).

\bibitem{Berker76}
A. Nihat Berker and M. Wortis, Phys. Rev. B {\bf 14}, 4946 (1976).

\bibitem{Jain80}
A. K. Jain and D. P. Landau, Phys. Rev. B {\bf 22}, 445 (1980).

\bibitem{Hasenbusch10}M. Hasenbusch,    arXiv:1004.4983v1.

\bibitem{Ozcelik08} V.O. {\"O}z\c{c}elik and A. N. Berker,
Phys. Rev. E {\bf 78}, 031104 (2008).

\bibitem{Fernandez08} L.A. Fern{\`a}ndez {\em et al.},
Phys. Rev. Lett. {\bf 100}, 057201 (2008).

\bibitem{Toldin09}  F.P. Toldin, A. Pelissetto and E. Vicari, J.  Stat. Phys. {\bf 135}, 1039 (2009).

\bibitem{Paoluzzi10} M. Paoluzzi, L. Leuzzi and A. Crisanti, Phys. Rev. Lett. {\bf}, (2010).

\bibitem{Hukushima96} K. Hukushima and K. Nemoto, J. Phys. Soc. Jpn. {\bf 65}, 1604 (1996).

\bibitem{Marinari98} E. Marinari, Adv. Computer Simul. {\bf 501}, 50 (1998).


%%%%%%%%%%%%%%%%%


%\bibitem{LPRR} L. Leuzzi {\em et al.},
% G. Parisi, F. Ricci-Tersenghi, and
%J. J. Ruiz-Lorenzo
%Phys. Rev. Lett. \textbf{101}, 107203 (2008).

%\bibitem{amit} D. J. Amit, \emph{Field Theory, the Renormalization
%Group, and Critical Phenomena} World Scientific.

%\bibitem{Allahverdyan}
%A.E. Allahverdyan and K.G. Petrosyan. Phys. Rev. Lett.{\bf 96}, 065701 (2006).


%%%%%%%%%%%%%%%%%
\bibitem{MPV86}M. M\`ezard, G. Parisi and M. A. Virasoro, \emph{Spin
glass theory and beyond} (Word Scientific, Singapore 1987)

\bibitem{Caracciolo93}S. Caracciolo {\em et al.}, Nucl. Phys. B403, 475
  (1993).

\bibitem{palassini} M. Palassini, S. Caracciolo,
Phys. Rev. Lett. \textbf{82}, 5128 (1999).

\bibitem{Kauz48} W. Kauzmann, Chem. Rev. {\bf 43}, 219 (1948).

\bibitem{Hill} T.L. Hill, {\em Thermodynamics of Small Systems}, Dover (2002).

\bibitem{Katzgraber_skew} R.S. Andrist, H.G. Katzgraber,
H. Bombin, M.A. Martin-Delgado, arXiv:1005.0777v1.

\bibitem{Riedel72} E.K. Riedel and F.J. Wegner,
Phys. Rev. Lett. \textbf{29}, 349 (1972).

\bibitem{ZZ} J. Zinn-Justin, {\em Quantum Field Theory and Critical
Phenomena}, Oxford University Press (Oxford, 1989).

\bibitem{Jorg06}
T. J\"org,  \emph{Phys. Rev. B} \textbf{73}, 224431 (2006).

\bibitem{HPV08} M. Hasenbusch, A. Pelissetto, E. Vicari,
\emph{J. Stat. Mech.} L02001, (2008).

\bibitem{BCPRB00} H. G. Ballesteros {\em et al.},
% A. Cruz, L. A. Fernandez,
%V. Martin-Mayor, J. Pech, J. J. Ruiz-Lorenzo, A. Tarancon, P. Tellez,
%C. L. Ullod, C. Ungil
Phys. Rev. B {\bf 62} (2000) 14237.

\bibitem{MPRPRB98} E. Marinari, G. Parisi and J. J. Ruiz-Lorenzo
Phys. Rev. B {\bf 58} (1998) 14852.

\bibitem{janus1} F. Belletti {\em et al.},
%, M. Cotallo, A. Cruz, L.A. Fernandez,
%A. Gordillo-Guerrero, M. Guidetti, A. Maiorano, F. Mantovani,
%E. Marinari, V. Martin-Mayor, A. Munoz Sudupe, D. Navarro, G. Parisi,
%S. Perez-Gaviro, J.J. Ruiz-Lorenzo, S.F. Schifano, D. Sciretti,
%A. Tarancon, R. Tripiccione, J.L. Velasco, D. Yllanes,
Phys. Rev. Lett. \textbf{101} 157201 (2008).

\bibitem{HTPV1} M. Hasenbusch,
 F. P. Toldin, A. Pellissetto, and
E. Vicari,
 Phys. Rev. B  \textbf{76}, 094402 (2007); {\em ibid.} 184202 (2007).

\bibitem{JK1} T. Jorg, H. G. Katzgraber, Phys. Rev. Lett. {\bf 101},
197205 (2008); Phys. Rev. B {\bf 77}, 214426
(2008).

\bibitem{sk} D. Sherrington, S. Kirkpatrick, \emph{Phys. Rev. Lett.}
\textbf{35}, 1792 (1975).

\bibitem{droplet} D. S. Fisher, D.A. Huse, Phys. Rev. Lett. {\bf 56},
1601 (1986).

\bibitem{tnt} F. Krzakala, O.C. Martin, Phys. Rev. Lett. {\bf 85},
3013 (2000).

\bibitem{rsb} G. Parisi, Phys. Lett. A {\bf 73} 203-205 (1979),
J. Phys. A {\bf 13}, L115 (1980), Phys. Rev. Lett. {\bf 50} 1946-1948
(1983).

\bibitem{KL} S. Kullback and R.A. Leibler, Ann. Math. Stat. {\bf 22},
79 (1951).
\bibitem{Contucci07}
P. Contucci {\em et al.}, Phys. Rev. Lett. 99, 057206 (2007).

\bibitem{Leuzzi08} L. Leuzzi, G. Parisi, F. Ricci-Tersenghi,
J.J. Ruiz-Lorenzo, Phys. Rev. Lett. {\bf 101}, 107203 (2008).

\bibitem{LeuzziNieuw07} L. Leuzzi and T.M. Nieuwenhuizen, {\em
Thermodynamic of the glassy state}, Taylor \& Francis (2007).

%\bibitem{footnote_last} For a system displaying both thermodynamic and random
%first order transitions see U. Ferrari, L. Leuzzi in preparation.



 \end{thebibliography}
\end{document}